\documentclass[a4paper]{jpconf}
\usepackage[english]{babel}
\usepackage[T1]{fontenc}
\usepackage{yhmath}
\usepackage{graphicx}
\graphicspath{{./}{Graphics/}}
\usepackage{booktabs}
\usepackage{subcaption}
\usepackage{array} 

\newcolumntype{L}[1]{>{\raggedright\let\newline\\\arraybackslash\hspace{0pt}}m{#1}}
\newcolumntype{C}[1]{>{\centering\let\newline\\\arraybackslash\hspace{0pt}}m{#1}}
\newcolumntype{R}[1]{>{\raggedleft\let\newline\\\arraybackslash\hspace{0pt}}m{#1}}

\usepackage{custom}
\usepackage{url}
\begin{document}
\title{Transient growth of wavelet-based resolvent modes in the buffer layer of wall-bounded turbulence}

\author{Eric Ballouz$^{1}$, Scott T. M. Dawson $^{2}$, H. Jane Bae$^{3}$}

\address{$^{1}$ Mechanical and Civil Engineering, California Institute of Technology, Pasadena CA 91125, USA}
\address{$^{2}$ Mechanical Materials, and Aerospace Engineering, Illinois Institute of Technology, Chicago IL 60616, USA}
\address{$^{3}$Graduate Aerospace Laboratories, California Institute of Technology, Pasadena CA 91125, USA}

\ead{eballouz@caltech.edu}

\begin{abstract}
In this work, we study the transient growth of the principal resolvent modes in the minimal flow unit using a reformulation of resolvent analysis in a time-localized wavelet basis.
We target the most energetic spatial wavenumbers for the minimal flow unit 
and obtain modes that are constant in the streamwise direction and once-periodic in the spanwise direction. 
The forcing modes are in the shape of streamwise rolls, though pulse-like in time, and the response modes are in the form of transiently growing streaks. 
We inject the principal transient forcing mode at different intensities into a simulation of the minimal flow unit and compare the resulting nonlinear response to the linear one. 
The peak energy amplification scales quadratically with the intensity of the injected mode, and this peak occurs roughly at the same time for all forcing intensities. 
However, the larger energy amplification intensifies the magnitude of the nonlinear terms, which play an important role in damping the energy growth and accelerating energy decay of the principal resolvent mode. 
We also observe that the damping effect of the nonlinearities is less prominent close to the wall. 
Finally, we find that the principal resolvent forcing mode is more effective than other structures at amplifying the streak energy in the turbulent minimal-flow unit. 
In addition to lending support to the claim that linear mechanisms are important to near-wall turbulence, this work identifies time scales for the nonlinear breakdown of linearly-generated streaks.

\end{abstract}

\section{Introduction}

Near-wall turbulence is believed to be organized into streamwise rolls, and alternating low- and high-speed streamwise streaks \cite{klebanoff1962three, kline1967structure, smithmetzler1983characteristics, 
blackwelder1979streamwise, johansson1987generation}. These coherent structures are well-documented and their characterization is the subject of multiple works \cite{bakewell1967viscous, landahl1980note, butler1993optimal, chernyshenko2005mechanism, delalamo2006linear}. Additionally, many studies point to a quasi-periodic cycle, wherein the streamwise streaks are amplified by streamwise vortices, meander, then break down, which subsequently regenerates new quasi-streamwise vortices \cite{kim1971production, hamilton1995regeneration, panton2001overview, jimenez2018coherent, smits2011high, adrian2007hairpin, robinson1991coherent}. The cycle, also known as the self-sustaining process, can be more clearly observed in a minimal flow unit \cite{jimenez1991minimal}, where the domain is artificially restricted in the streamwise and spanwise directions in order to exclude the dynamics of the outer region of the channel. Despite the restriction, the study of the minimal flow unit reveals that the near-wall cycle can self-sustain even when motions at larger scales are inhibited \cite{jimenez1999autonomous}.

Although nonlinear mechanisms play a role in the self-sustaining process, a lot of attention has been given to linear mechanisms and instabilities as the drivers of this process \cite{panton2001overview, lozano2021cause}. 
One example is the Orr mechanism \cite{jimenez2013linear, orr1907stability}, in which the mean shear near the wall tilts velocity perturbations forward in the streamwise direction and stretches vertical scales, intensifying the wall-normal velocity perturbations. 
Another example is lift-up \cite{hwang2010self}, which occurs when wall-normal velocity perturbations transport slow-moving fluid near the wall away into the faster flow field farther away from the wall.
Works like Del Alamo and Jim\'enez \cite{delalamo2006linear} and Pujals \emph{et al.} \cite{pujals2009note} show that, even after removing the nonlinear term from the perturbation equations, linear transient growth via the mean shear generates the dominant (streaky) structures in wall-bounded turbulence. The linearized system additionally accounts for much of the energy spectra and reproduces the self-similar profile in the logarithmic region. 
Similarly, Lozano-Dur\'an \emph{et al.} \cite{lozano2021cause} show through numerical experiments that the minimal flow unit can sustain turbulence without the nonlinear feedback between the velocity fluctuations and the mean velocity profile, except when the Orr-mechanism or push-over (momentum transfer from the spanwise mean shear into the streamwise velocity perturbation) are suppressed. The authors thus argue for linear transient growth as a prominent mechanism for transferring energy from the mean flow to turbulent
fluctuations.

The linear amplification process linking streamwise vortices and streamwise streaks has also been fruitfully studied through the lens of resolvent analysis \cite{mckeon2010critical,moarref2014foundation,mckeon2017engine}.
In resolvent analysis, the Navier-Stokes equations are reframed as a linear dynamical system for the velocity fluctuations, and the nonlinear term as external forcing acting on this system. The goal is then to solve for the spatial structure of the (nonlinear) forcing that generates the response (velocity) with the largest linear energy amplification. 
In the context of wall-bounded turbulent flows, the method is successful at identifying streamwise rolls as the most perturbing structures, and streamwise streaks as the most amplified structures. 
Furthermore, Bae \emph{et al.} \cite{bae2021nonlinear} demonstrate the role of these linearly-identified resolvent modes in transferring energy to coherent near-wall turbulent perturbations, even within a fully nonlinear turbulent flow. Their study shows that projecting out the contribution of the leading resolvent forcing mode from the nonlinear term at every time step interrupts the streak-regeneration process and greatly suppresses buffer layer turbulence.  

In resolvent analysis, the resolvent operator is  constructed from the linearized Navier-Stokes equations that are Fourier transformed in the homogeneous spatial directions and in time. For turbulent channel flow, this means the domain is usually assumed to be periodic in the streamwise and spanwise directions and statistically stationary in time \cite{mckeon2010critical, mckeon2017engine, moarref2014foundation, bae2018}. These resolvent modes, which are Fourier modes in time, lack transient linear growth information. In light of the importance of transient linear growth to wall-bounded turbulence, the aim of this work is to reformulate resolvent analysis in a time-localized basis, such as one constructed with wavelets rather than Fourier modes \cite{ballouz2023wavelet}, and to further examine the interactions of these transient resolvent modes with the fully nonlinear turbulent flow in the minimal flow unit.

This paper is organized as follows. In \S\ref{resolvent}, we compute resolvent modes in a time-localized wavelet basis as in Ballouz \emph{et al.} \cite{ballouz2023wavelet}, and constrain the forcing to a wavelet-shaped pulse. This formulation yields a forcing mode in the shape of streamwise rolls that is compactly supported in time, in addition to an optimal streak-like response that grows transiently before decaying. The justification for the choice of spatial wavenumbers and wavelets is given in \S\ref{parameters}. We then solve the fully nonlinear forced Navier-Stokes equations for the minimal flow unit at $Re_\tau = 186$, using the time-localized wavelet-based resolvent forcing mode as our forcing term. This step is detailed in \S\ref{dns}. We track the evolution of this resolvent forcing mode as it generates and amplifies streamwise streaks, and compute relevant turbulent statistics, which we present in \S\ref{results}. Concluding remarks are given in \S\ref{conclusion}.

\section{Methods}


In this work, we consider the flow in the minimal flow unit of size $L_x \times L_y \times L_z = 1.72\delta \times 2\delta \times 0.86 \delta$, where $\delta$ is the channel half-height, and $x$, $y$ and $z$ are the streamwise, wall-normal and spanwise directions, respectively. We denote the velocity fluctuation field by $\boldsymbol u = [u, v, w]^T$, where $u$, $v$ and $w$ are the streamwise, wall-normal, and spanwise components respectively. The system is characterized by the friction Reynolds number $Re_\tau = \delta u_\tau / \nu \approx 186$, where $u_\tau$ is the friction velocity, and $\nu$ is the kinematic viscosity. The flow is periodic in the streamwise and spanwise directions, and the no-slip and no-penetration conditions hold at the walls of the channel.

For the direct numerical simulations (DNS) in this work, we discretize the streamwise and spanwise directions uniformly using $N_x = N_z = 32$ grid points, which results in streamwise and spanwise grid spacings of $\Delta x^+ \approx 10$ and $\Delta z^+ \approx 5$. In the wall-normal direction, we use a Chebyshev grid of size $N_y =  128$, which results in a wall-normal spacing of $\min(\Delta y^+) \approx 0.17$ near the wall and $\max(\Delta y^+) \approx 7.6$ at the centerline. Here, the superscript $+$ denotes wall units normalized with $u_\tau$ and $\nu$. We discretize the incompressible Navier-Stokes equations with a staggered, second-order-accurate, central finite difference method in space \cite{orlandi2000fluid}, and a fractional step method is used to compute pressure \cite{kim1985application}. Time-advancement is performed with an explicit third-order-accurate Runge-Kutta method \cite{wray1990minimal}. The DNS code has been validated in previous studies of turbulent channel flows \cite{lozano2016turbulent, bae2018turbulence, bae2019dynamic}. Using this discretization, we obtain a mean streamwise velocity profile $U(y)$ by averaging DNS results of the unforced system in the homogeneous directions and time. This profile is used in the subsequent sections.


\subsection{Spatio-temporal resolvent modes}\label{resolvent}

We first compute the time-localized resolvent forcing modes and their corresponding transient responses for the minimal flow unit. The incompressible Navier-Stokes equations for the velocity fluctuations about the mean turbulent flow field $\boldsymbol{U}(y) = (U(y), 0, 0)$ are Fourier-transformed in the $x$- and $z$-directions, and wavelet-transformed in time, \emph{i.e.}
\begin{equation}\label{NS}
    \widetilde \partial_t \boldsymbol{\tilde u} = -(\boldsymbol{U} \cdot \tilde \nabla) \boldsymbol{\tilde u} - (\boldsymbol{\tilde u} \cdot \tilde \nabla) \boldsymbol{\tilde U} - \frac{1}{\rho}\tilde \nabla \tilde p + \nu \tilde \Delta \boldsymbol{\tilde u} + \boldsymbol{\tilde f}, \; \; \tilde \nabla \cdot \boldsymbol{\tilde u} = 0,
\end{equation}
where $(\tilde \cdot)$ denotes a Fourier transform for a given streamwise and spanwise wavenumber pair $(k_x, k_z)$ composed with a wavelet transform in time, $\rho$ is density, and $p$ is the pressure. 
These transformed quantities are functions of wall-normal position $y$, and the wavelet scale and shift parameters $\alpha$ and $\beta$ 
, which respectively determine the time interval and frequency range of the wavelet mode onto which we project our physical quantities
\cite{ballouz2023wavelet}. The nonlinear term, denoted by $\boldsymbol{\tilde f}$, is assumed to act as an exogenous forcing function on the velocity field. The modified spatial derivative operators are given by $\tilde \nabla = [\text{i}k_x, \partial_y, \text{i}k_z]^T$, where $\text{i}^2 = -1$, and $\tilde \Delta = -k_x^2 + \partial_{yy} - k_z^2$, and the modified time derivative operator is defined as $\tilde \partial_t = \mathcal W \circ\partial_t \circ \mathcal W^{-1}$, where $\mathcal W$ is the discrete wavelet transform. The transform $\mathcal W$ projects an arbitrary function onto a set of shifted and rescaled wavelet and scaling functions \cite{mallat2001wavelet,najmi2012wavelets}. The choice of wavelet-scaling-function pair determines the properties of the transform, most notably its sparsity, which depends on the compactness of the chosen functions, and whether it is unitary, which depends on whether the wavelet and scaling function bases are orthonormal. For a discussion on the choice of wavelet transform, see \S\ref{parameters}.

We can rearrange (\ref{NS}) as
\begin{equation}\label{inout}
    \begin{bmatrix}
        \boldsymbol{\tilde u}(y, \alpha, \beta) \\
        \tilde p(y, \alpha, \beta)
    \end{bmatrix} = \widetilde{\mathcal H}^{(k_x, k_z)} \begin{bmatrix}
        \boldsymbol{\tilde f}(y, \alpha, \beta) \\ 0
    \end{bmatrix},
\end{equation}
where the linear operator $\widetilde{\mathcal H}^{(k_x, k_z)}$ is the resolvent operator in this formulation. In the setting of a discretized spatial and temporal domain, $\widetilde{\mathcal H}^{(k_x, k_z)}$ is a $(4N_y N_t) \times (4 N_y N_t)$ matrix, where $N_t$ is the resolution in time. Indeed, this resolvent formulation targets all temporal basis functions, rather than one as in traditional resolvent analysis. The factor $4$ is due to the fact that the left-hand side of equation \eqref{inout} contains four variables, $\boldsymbol{\tilde u}$, and $\tilde p$. 
Note the dependence on the spatial parameters $(k_x, k_z)$, which must be chosen to define the Fourier transform in the homogeneous directions. 
To discretize the wall-normal direction, we use the same Chebyshev grid as in the DNS. The spatial derivatives in $y$ are similarly staggered second-order-accurate central finite differences, and the Fourier spatial derivatives in the streamwise and spanwise directions are computed with the modified wavenumbers that match the choice of differentiation scheme used in the DNS. We choose a second-order-accurate central finite difference matrix for the time derivative operator. A total time of $T = 22 \; \delta / u_\tau$ is used to allow the resolvent response modes to decay to zero, allowing us to use periodic boundary conditions for the time derivative matrix. Though not shown, increasing the order of the finite difference operator does not strongly affect the resolvent modes. 

We introduce the additional step of constraining the forcing along a wavelet-shaped pulse of any desired scale $\alpha$ and shift $\beta$ using a temporal windowing matrix $\mathcal B$ \cite{jeun2016input, kojima2020, ballouz2023wavelet}. 
We then take the singular value decomposition (SVD) of the combined operator 
\begin{equation}
    \widetilde{\mathcal H}^{(k_x, k_z)} \mathcal B = \sum_{j=1}^\infty \sigma_j \boldsymbol{\tilde \psi_j}(y, \alpha, \beta) \otimes \boldsymbol{\tilde \phi_j}(y, \alpha, \beta),
\end{equation}
where we index the singular values $\{\sigma_i\}_{i=1}^{\infty}$ such that $\sigma_1 \geq \sigma_2 \geq ... \geq 0$. The right and left singular vectors $\{\boldsymbol{\tilde \phi_j}\}_{j=1}^{\infty}$ and $\{\boldsymbol{\tilde \psi_j}\}_{j=1}^{\infty}$ respectively define orthonormal bases for the spaces containing the nonlinear term (forcing) and the velocity and pressure fluctuations (response). For the SVD, we choose the inner product to be the kinetic energy seminorm, which we define for an arbitrary vector  $\boldsymbol{\tilde b} =  [\tilde b_u, \tilde b_v, \tilde b_w, \tilde b_p]^T$ to be
\begin{equation}
    \|\boldsymbol{\tilde b} \|^2 = \frac{u_\tau}{\delta} \frac{1}{L_x(2\delta)L_z}\int_0^{L_x}\int_0^{2\delta} \int_{0}^{L_z}\int_{-\infty}^{+\infty} (|b_u|^2 + |b_v|^2 + |b_w|^2 )\,\mathrm{d}t\,\mathrm{d}z\,\mathrm{d}y \,\mathrm{d}x,
\end{equation}
where $\boldsymbol{b} = [b_u, b_v, b_w, b_p]^T$ is the inverse transform of $\boldsymbol{\tilde b}$. 
The kinetic energy amplification factor is given by the square of the singular values. The forcing modes are therefore ordered decreasingly according to the integrated kinetic energy amplification they undergo when acted on by $\widetilde{\mathcal H}^{(k_x, k_z)} \mathcal B $, and the response modes are the corresponding amplified coherent structures arising from this action. 
Thus, $\boldsymbol{\tilde \phi_1}= [\tilde \phi_{1, u}, \; \tilde \phi_{1, v}, \; \tilde \phi_{1, w}, \;0]^T$ generates the largest linear energy amplification via the windowed resolvent operator, and $\sigma_1 \boldsymbol{\tilde \psi_1} = \sigma_1 [\tilde \psi_{1, u}, \; \tilde \psi_{1, v}, \; \tilde \psi_{1, w}, \; \tilde \psi_{1, p}]^T$ is the resulting optimally-amplified transient velocity and pressure fluctuation.
For all $j$, $\boldsymbol{\phi_j}$, the inverse transform of $\boldsymbol{\tilde \phi_j}$, is shaped in time according to the wavelet or scaling function chosen by $\mathcal B$. As described in Bae \emph{et al.} (2021), the resolvent modes occur in equivalent pairs of equal singular values due to the symmetry of the channel geometry. For the numerical experiments in \S\ref{dns}, we linearly combine the two equivalent forcing modes (e.g. $\boldsymbol{\tilde \phi_1}$ and $\boldsymbol{\tilde \phi_2}$) to isolate the support of the forcing mode to the bottom half of the channel. Thus, upon injecting this mode into the DNS of the forced Navier-Stokes equations, only the bottom of half of the channel is subject to the forcing, allowing us to use the top half as a control system \cite{bae2021nonlinear}. Henceforth, $\boldsymbol{\tilde \phi_1}$ will refer to the linear combination of the first two equivalent forcing modes that limits the forcing to the bottom-half of the channel. $\boldsymbol{\tilde \phi_3}$ is defined similarly with regards to the third and fourth forcing modes.

\subsection{Choice of spatial and temporal scales} \label{parameters} 
As in Bae \emph{et al.} (2021), we choose to study the modes given by wavenumbers of $k_x L_x/(2\pi) = 0$ and $k_z L_z/(2\pi) = 1$, which account for a peak in the spectral energy content for the minimal flow unit at a wall-normal height of $y^+ = 15$. This is consistent with the idea that the minimal flow unit is the smallest channel capable of sustaining turbulent streaks; the single streak thus appears to stretch the entire length of the domain hence the streamwise wavenumber of $0$. Similarly, the domain is only large enough to contain one low- and high-speed streak pair in the spanwise direction, which yields the spanwise wavenumber of $1$.

\begin{figure} 
    \centering
    \begin{subfigure}{0.44\textwidth}
        \includegraphics[width=\linewidth]{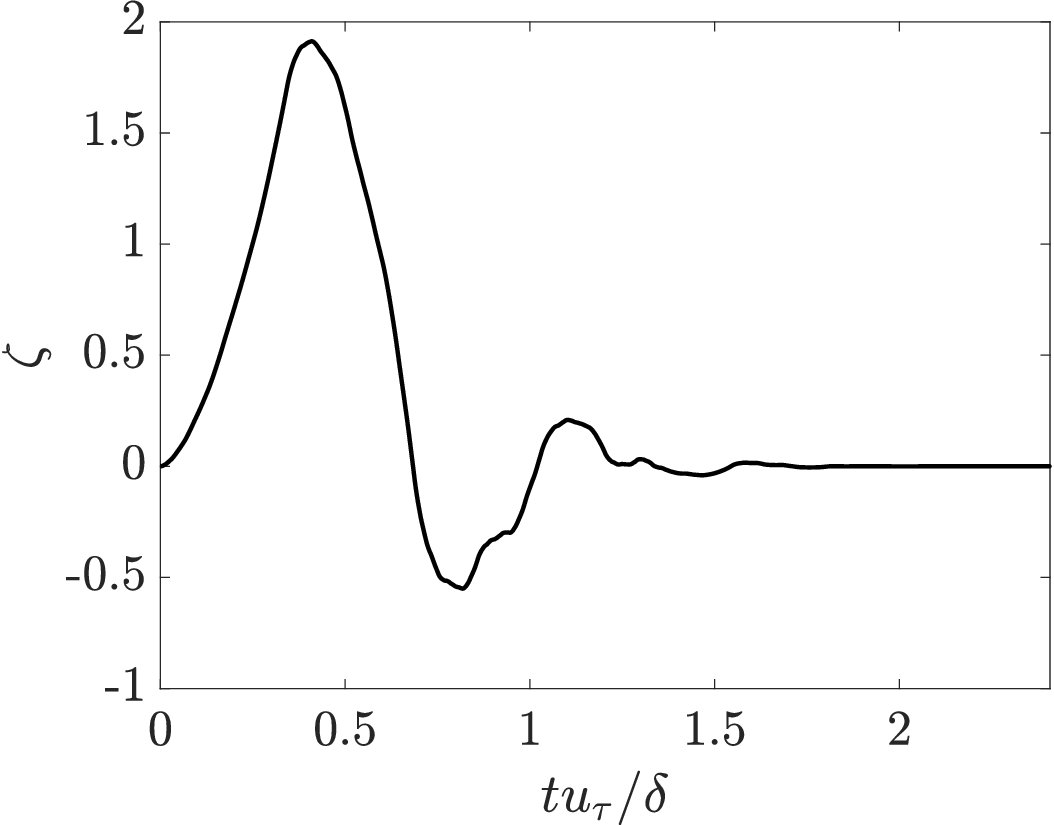}
        \caption{}
    \end{subfigure}
    \begin{subfigure}{0.46\textwidth}
        \includegraphics[width=\linewidth]{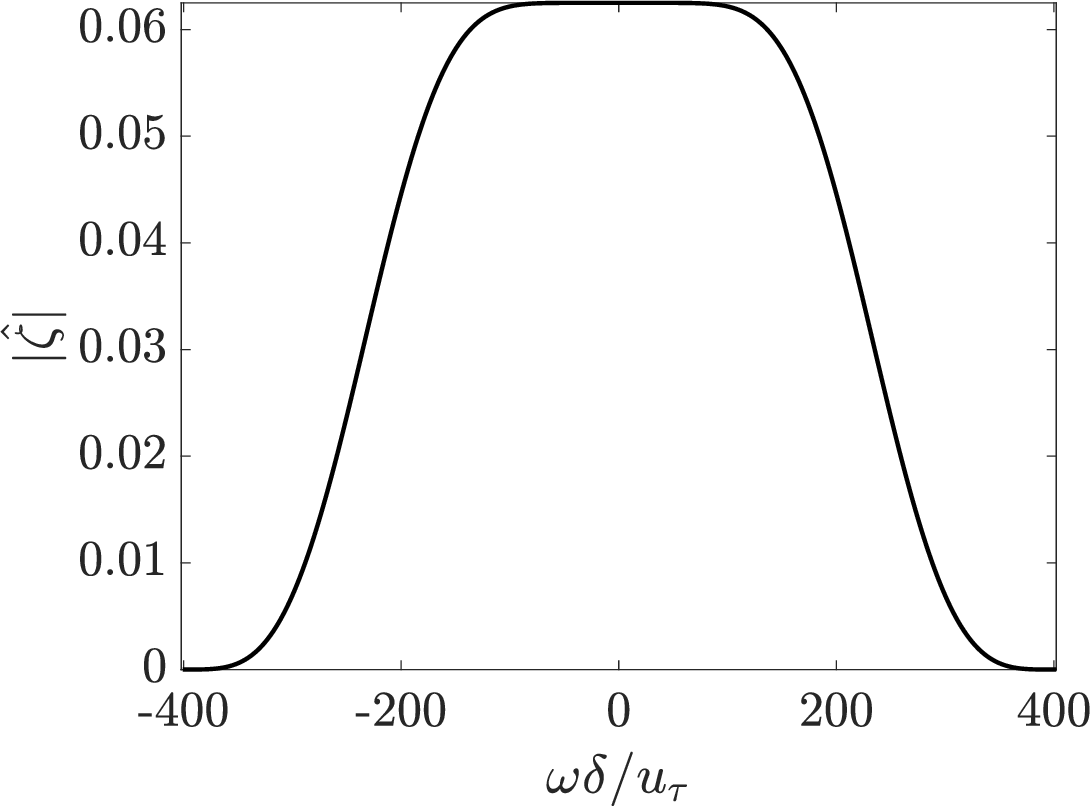}
        \caption{}
    \end{subfigure}
    \caption{Daubechies-8 scaling function $\zeta$ in time (a) and frequency (b) domain.
    }
    \label{Daub8}
\end{figure}

Traditional resolvent analysis in which the Navier-Stokes equations are Fourier-transformed in time reveals that a temporal frequency of $\omega = 0$ corresponds to the largest linear amplification \cite{bae2021nonlinear}. However, since we constrain the forcing term to a pulse that is compactly-supported in time, resolvent analysis will not be able to target a single frequency, but will capture a wide range of frequencies. A trade-off exists between time and frequency localization, and the more precision we require in one domain, the less we preserve in the other \cite{mallat2001wavelet, najmi2012wavelets}. 
In this work, we use a single-level Daubechies-8 wavelet transform \cite{daubechies1992ten}. Using $\mathcal B$, we constrain the forcing term to the desired Daubechies-8 scaling function of arbitrary shift $\beta$ (\emph{i.e.} centered at an arbitrary time) denoted by $\zeta(t)$, since it is compactly-supported in time and its Fourier-spectrum is a quasi band-pass filter that encompasses the target frequency $\omega = 0$ resulting in the largest linear amplification via the traditional resolvent operator for the minimal flow unit. The scaling function satisfies $\left ( \int_{-\infty}^{+\infty} |\zeta(t)|^2 \mathrm{d}t \right ) \; u_\tau / \delta = 1$. The chosen scaling function is shown along with its Fourier spectrum in figure \ref{Daub8}. We note that the simulations detailed in \S\ref{dns} resolve temporal wavenumbers up to $\omega \delta / u_\tau \approx 55,000$.

\begin{figure}
    \centering
    \begin{subfigure}{0.47\textwidth}
        \includegraphics[width=\linewidth]{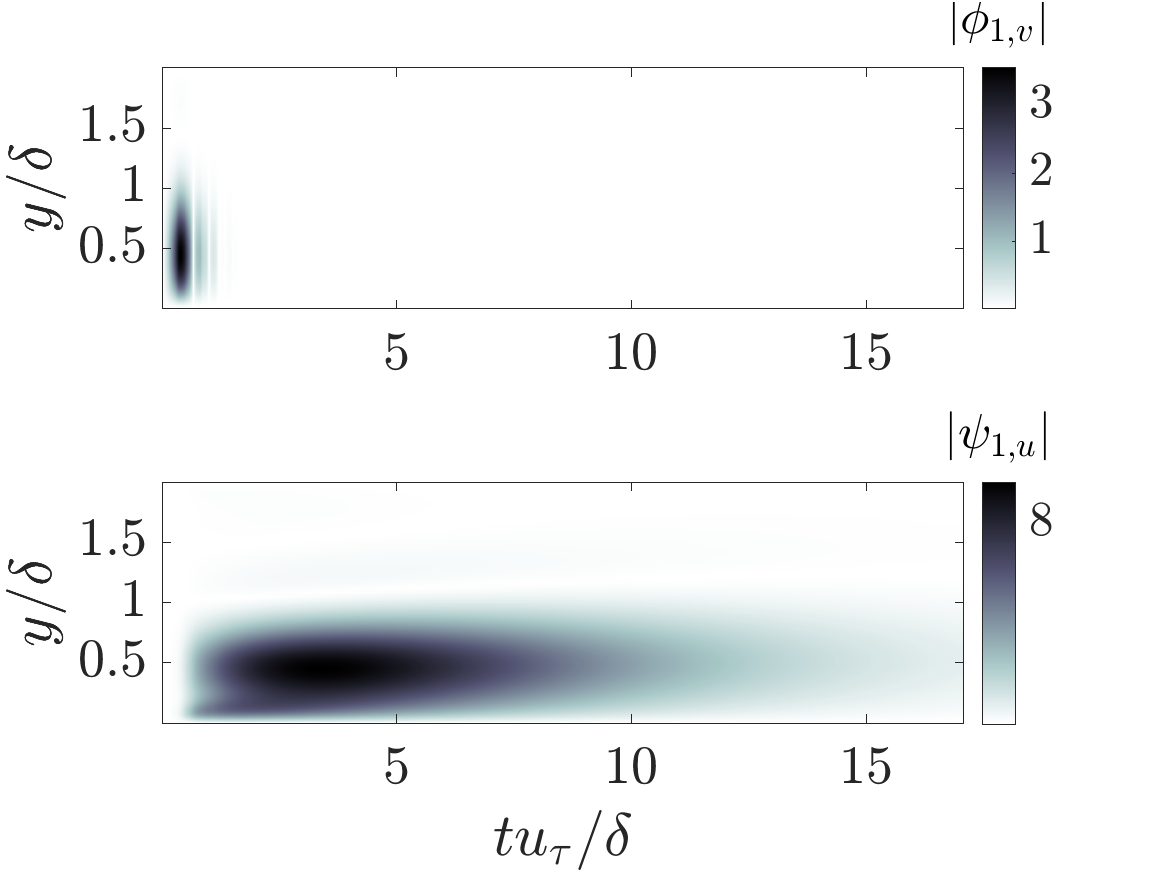}
    \caption{}
    \end{subfigure}
    \begin{subfigure}{0.43\textwidth}
        \includegraphics[width=\linewidth]{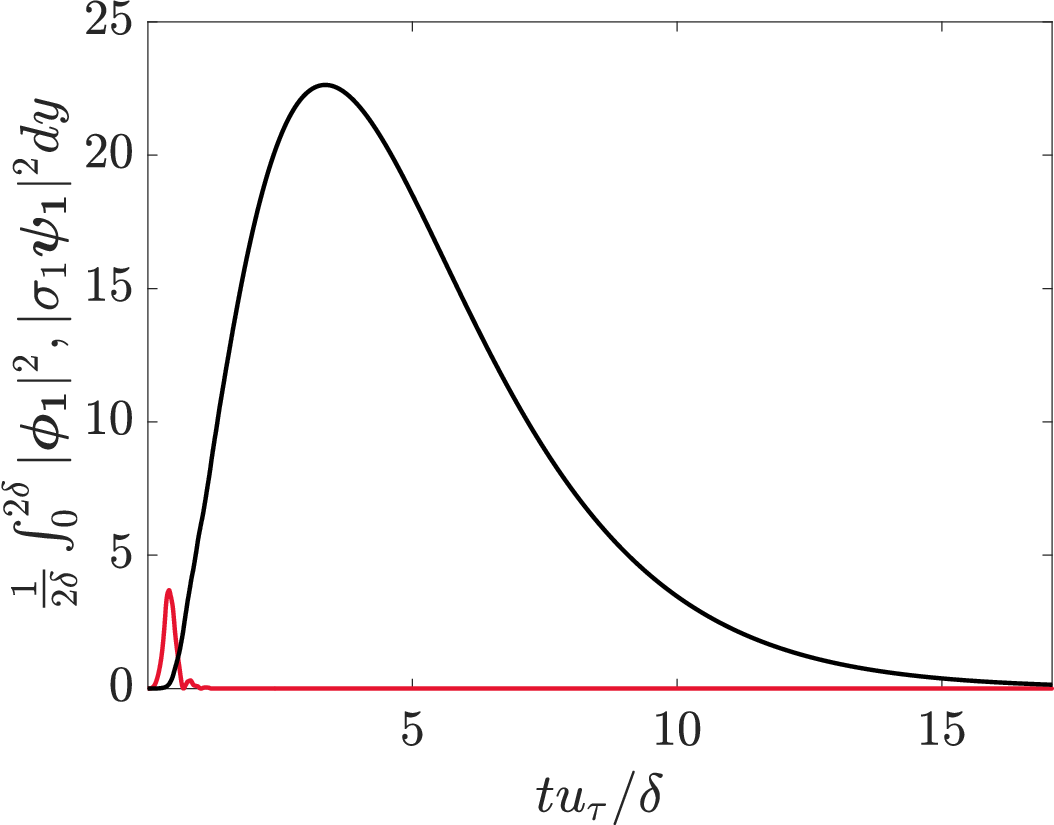}
    \caption{}
    \end{subfigure}
        \begin{subfigure}{0.45\textwidth}
        \includegraphics[width=\linewidth]{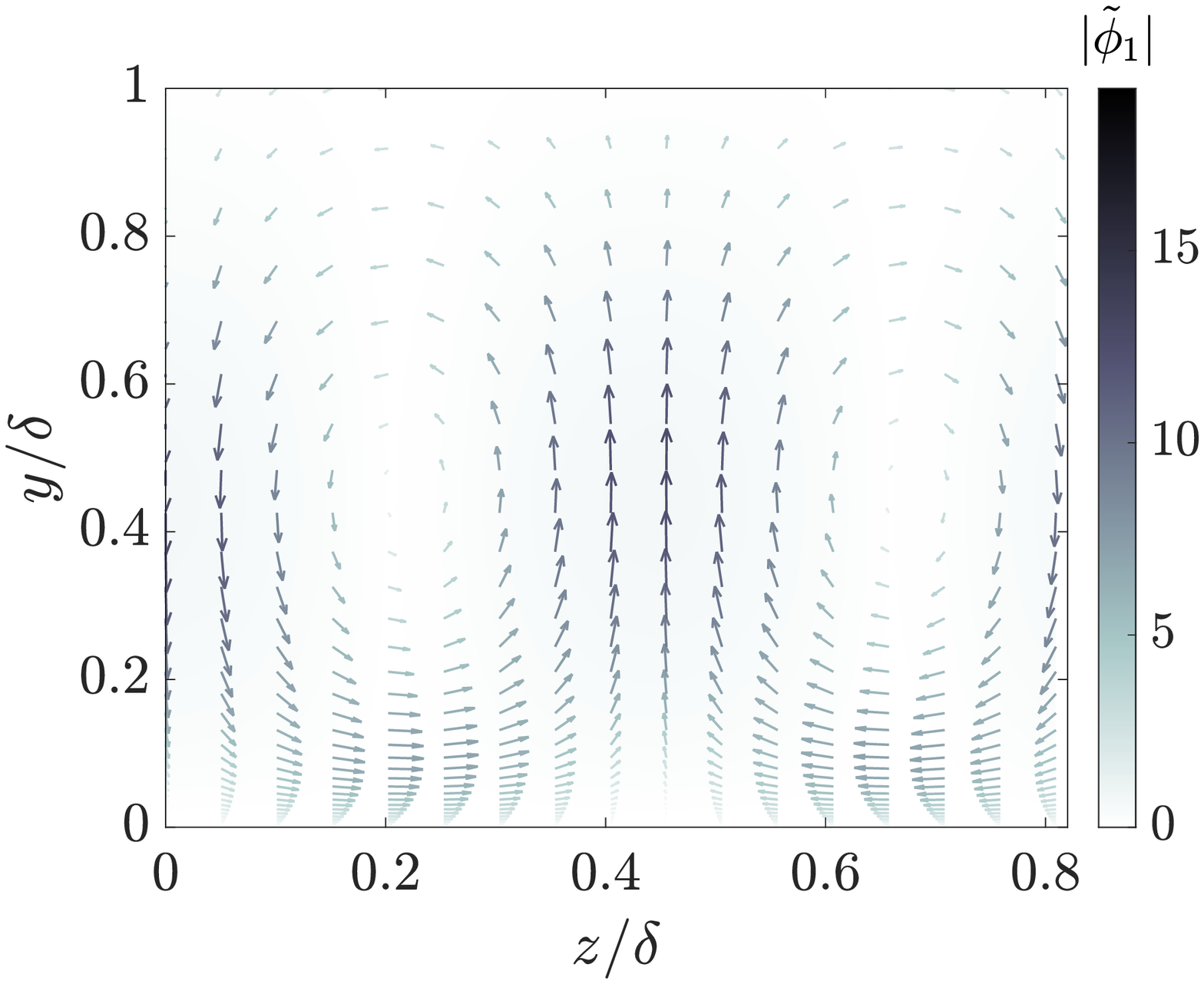}
    \caption{}
    \end{subfigure}
    \begin{subfigure}{0.45\textwidth}
        \includegraphics[width=\linewidth]{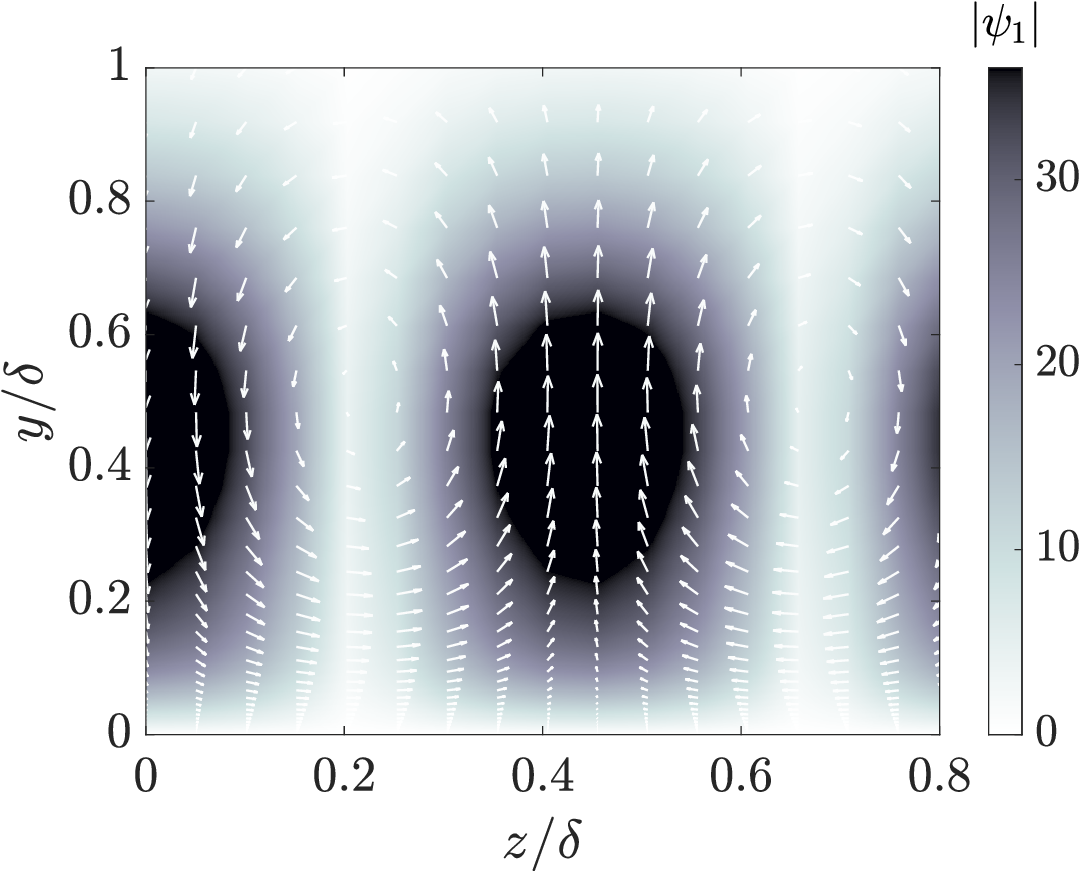}
    \caption{}
    \end{subfigure}
    \caption{(a) Magnitudes of the wall-normal component of the principal forcing mode (top) and the streamwise component of the principal response mode (bottom). (b) Integrated energy of the principal forcing (red) and response (black) modes. (c) Principal forcing mode shown at peak amplitude ($t u_\tau /\delta \approx 0.40)$. (d) Principal response modes at peak amplitude ($t u_\tau /\delta \approx 2.41)$. In (c,d), the contours represent the streamwise magnitude; the arrows, which show the direction of the cross flow components, are colored according to their magnitudes $\sqrt{| \phi_{1,v}|^2 + | \phi_{1,w}|^2}$ in (c), or $\sqrt{| \psi_{1,v}|^2 + |\psi_{1,w}|^2}$ in (d).}
    \label{linearmodes}
\end{figure}

The obtained resolvent modes are shown in figure \ref{linearmodes}. Notably, the response modes exhibit transient energy growth and decay as seen in figure \ref{linearmodes}(a,b). The inverse transforms of the modes are shown in figure \ref{linearmodes}(c,d). The optimal transient nonlinear forcing mode appears in the shape of streamwise rolls, and the optimal velocity fluctuation response appears as predominantly streamwise streaks with alternating signs of the same magnitude. This supports the extensively-examined claim that streamwise streaks can be linearly generated by a linear lift-up mechanism, whereby slower moving fluid close to the wall is swept upwards into the faster moving mean flow farther away from the wall. The streak-shaped response mode grows in intensity before fading, showcasing the transient growth characteristic of the linearized Navier-Stokes equations for this system.

\subsection{Forced direct numerical simulations}\label{dns}

We first perform a DNS with a fixed mean flow given by $\boldsymbol{U} = [U(y), 0, 0]^T$, which was used to calculate the resolvent modes. From this simulation, we obtain snapshots with a frozen mean profile and use these to initialize the ensemble of the resolvent-forced simulations. 
Freezing the mean profile ensures that the DNS mean profile matches the one used to compute the resolvent modes for all time. We do this by initializing the flow to have the desired mean streamwise profile of $U$, then by removing the steady-state contribution of the right-hand side of the Navier-Stokes equations. The initial snapshots from the fixed-mean simulation are separated by $\Delta t \approx 17.5\; \delta / u_\tau$, and amount to an ensemble size of approximately 990.

For each resolvent-forced simulation, we have a corresponding unforced fixed-mean simulation with the same initial condition. We denote the velocity fluctuations for the unforced simulations by $\boldsymbol{u}_0$, and the nonlinear terms in the momentum equations by $\boldsymbol{g}_0$.
For the resolvent-forced DNS, we introduce a forcing term proportional to $\boldsymbol{ \phi_1} $, the inverse transform of the principal resolvent forcing mode $ \boldsymbol{\tilde{\phi_1}} $.  
The forcing, normalized so that $\| \boldsymbol{\tilde{\phi_1}}\|^2 = 1$, 
is scaled by a complex constant $\kappa$ with magnitude
\begin{equation}
    |\kappa| := \gamma  \left (\frac{\delta}{u_\tau^3} \frac{1}{L_x (2\delta)L_z}\int_{-\infty}^{+\infty}\int_0^{L_x}\int_0^{2\delta}\int_0^{L_z} \left \vert \frac{\partial \boldsymbol{{u}_0}}{\partial t}\right \vert^2_{t=0} \zeta(t)^2\mathrm{d}z \, \mathrm{d}y \, \mathrm{d}x\, \mathrm{d}t \right )^{1/2},
\end{equation} 
where $\gamma \in \{ 1\%, 2\%, 5\%, 10\%\}$ such that the resolvent forcing mode is increasing the initial energy of the right-hand side by $\gamma \%$. 
Thus, $|\kappa|^2$ determines the integrated energy injected into the system by the forcing. We choose $\angle \kappa = \angle \langle \partial_t \boldsymbol{\hat u}_0^{(0, 1)}, \phi_1 \rangle$, so the forcing mode is in phase with the right-hand side of the unforced flow-field. 
To test the optimality of $\boldsymbol{\phi_1}$ at forcing the turbulent channel, we repeat the case with $\gamma = 5\%$ using $\boldsymbol{\phi_3}$ and a randomly generated term $\boldsymbol{\phi}_{\text{\bf{rand}}} = \hat {\boldsymbol{\phi}}_{\text{\bf{rand}}}(y) \zeta(t)$, which we normalize and scale the same way. 
As in the unforced case, the mean profile of the forced simulation is fixed, and we run the forced DNS for a total time of $T = 5.69 \; \delta / u_\tau$.

\section{Results and discussion} \label{results}
In this section, we define the deviation operator $\Delta$ as the difference between the forced and unforced simulations, e.g., $\Delta \hat{u}^{(0,1)} = \hat u^{(0,1)} - \hat u_0^{(0,1)}$. 
The velocity deviation field is then given by 
$\boldsymbol{\hat q} = [\Delta \hat u^{(0,1)}, \Delta \hat v^{(0,1)}, \Delta \hat w^{(0,1)}]^T$, and the deviation in the nonlinear terms $\boldsymbol{\hat f} = \Delta \boldsymbol{\hat g}^{(0,1)}$, where $\boldsymbol{\hat{g}}$ is the nonlinear terms in the Navier-Stokes equations. 
We define the instantaneous streak energy as 
\begin{equation}
    \hat E_u^{(0,1)}(t) = \frac{1}{2\delta} \int_0^{2\delta} \frac{|\hat u^{(0,1)}|^2}{2} \mathrm{d}y,
\end{equation}
and the deviation in total kinetic energy as $\Delta  \hat E^{(0,1)}(t)$.
Finally, we denote the ensemble average by $\overline{ (\cdot) }$.

\subsection{Transient energy growth and decay of streaks in the forced DNS}\label{transientEnergyGrowth}

\begin{figure}
    \centering
    \begin{subfigure}{0.44\textwidth}
    \includegraphics[width=\linewidth]{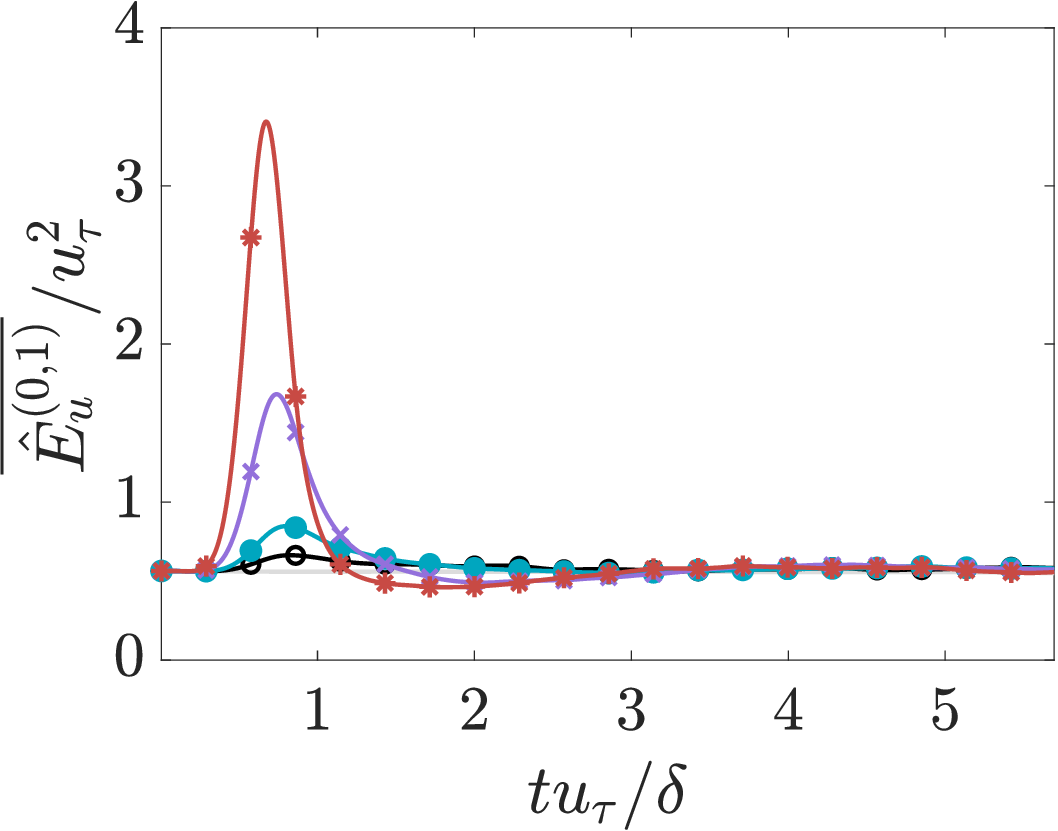}
    \caption{}
    \end{subfigure}
    \begin{subfigure}{0.45\textwidth}
    \includegraphics[width=\linewidth]{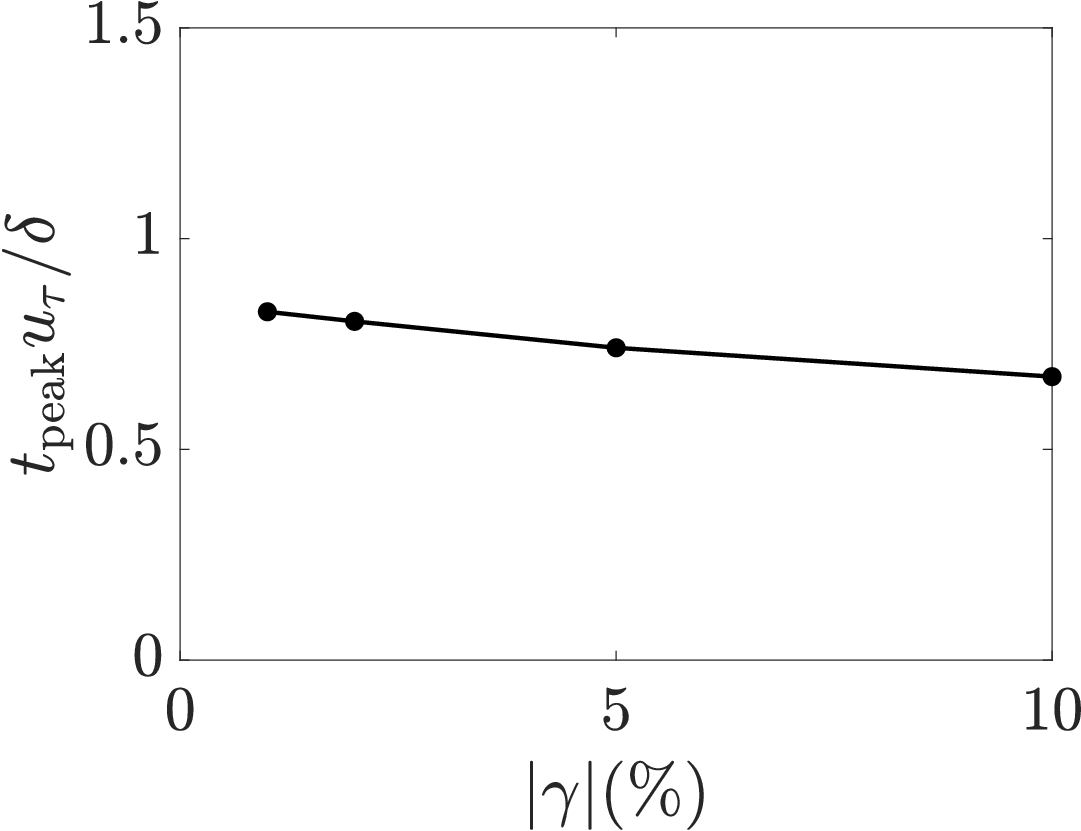}
        \caption{}
    \end{subfigure}
    \begin{subfigure}{0.45\textwidth}
       \includegraphics[width=\linewidth]{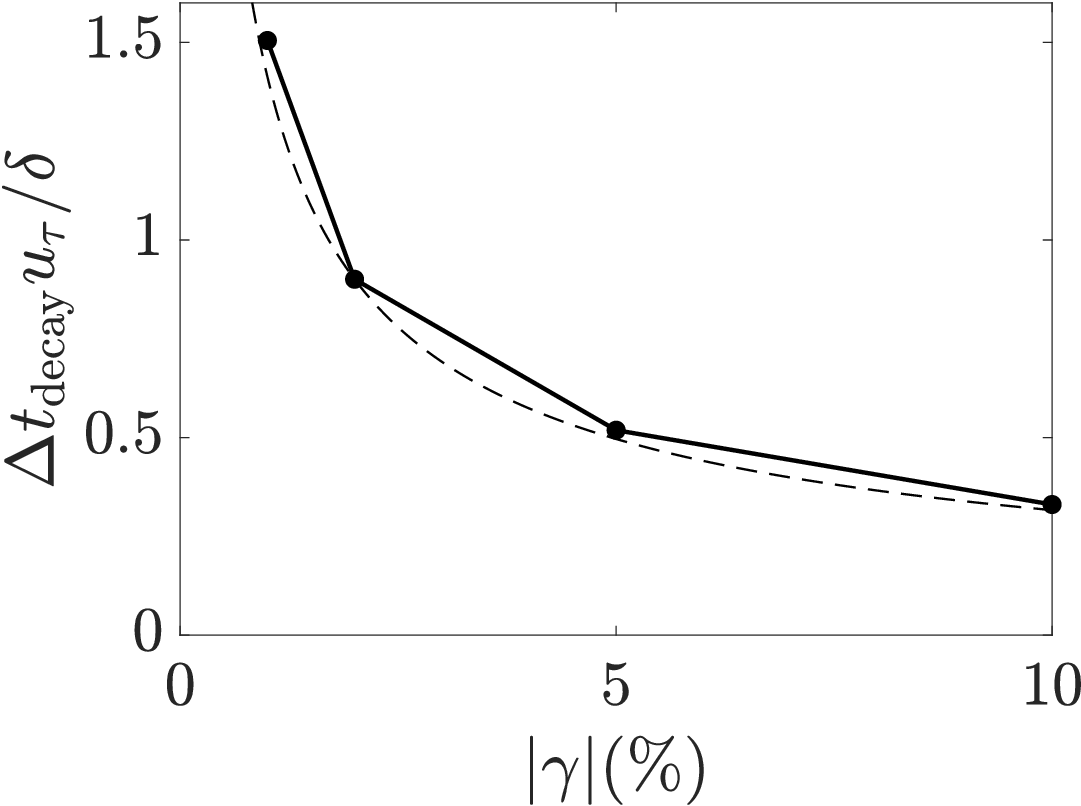}
        \caption{}
    \end{subfigure}
        \begin{subfigure}{0.45\textwidth}
        \includegraphics[width=\linewidth]{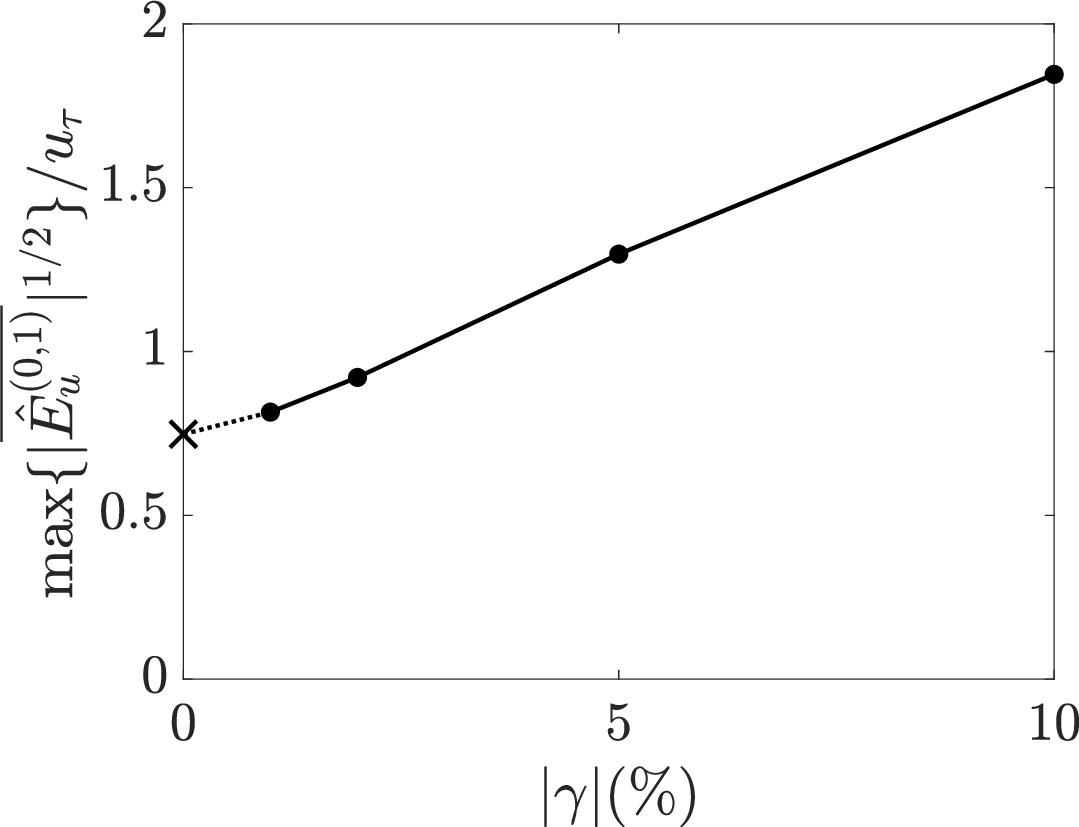}
        \caption{}
    \end{subfigure}
    \caption{Average streak energy as a function of time (a), peak time (b), decay time (c), and streak energy peaks (d) as a function of $\gamma$. The cases plotted in (a) are $\gamma = 1\%$ (black $\circ$), $\gamma = 2\%$ (cyan $\bullet$), $\gamma = 5\%$ (purple $\times$), and $\gamma = 10\%$ (red $*$). The dashed line in (c) is $|\gamma|^{-0.65}$. In (d), $\times$ denotes the streak energy in the unforced case.}
    \label{transientStats}
\end{figure}

Figure \ref{transientStats}(a) shows the streak energy contained in the $(0, 1)$-mode as a function of time, for different resolvent-forcing amplitudes. For all cases, the energies grow and peak before decaying. The stronger the forcing, the faster the streak energy's growth rate, and the faster its decay. We note that the differences in decay rate are more dramatic compared to those in growth rate across the different forcing amplitude cases. The peak times, $t_{\text{peak}}$, defined as the times at which the energies reach their maxima, decrease slightly with forcing amplitude, but are relatively constant compared to the decay times, $\Delta t_{\text{decay}} = t_{10\%} - t_{\text{peak}}$, which we define as the time it takes for the energy to reduce from the peak to $10\%$ of its peak (figures \ref{transientStats}b, \ref{transientStats}c). 
Indeed, the decay time varies widely and scales as $\Delta t_\text{decay} \sim |\gamma|^{-0.65}$. 
We note that all fully-coupled simulations decay significantly faster ($t \approx 1 \delta / u_\tau$) than the linear response ($t \approx 15 \delta / u_\tau$). In addition to energy growth and decay, we observe that the peak energy scales quadratically with the forcing amplitudes (figure \ref{transientStats}d). 



\begin{figure}
    \centering
    \includegraphics[width=0.5\linewidth]{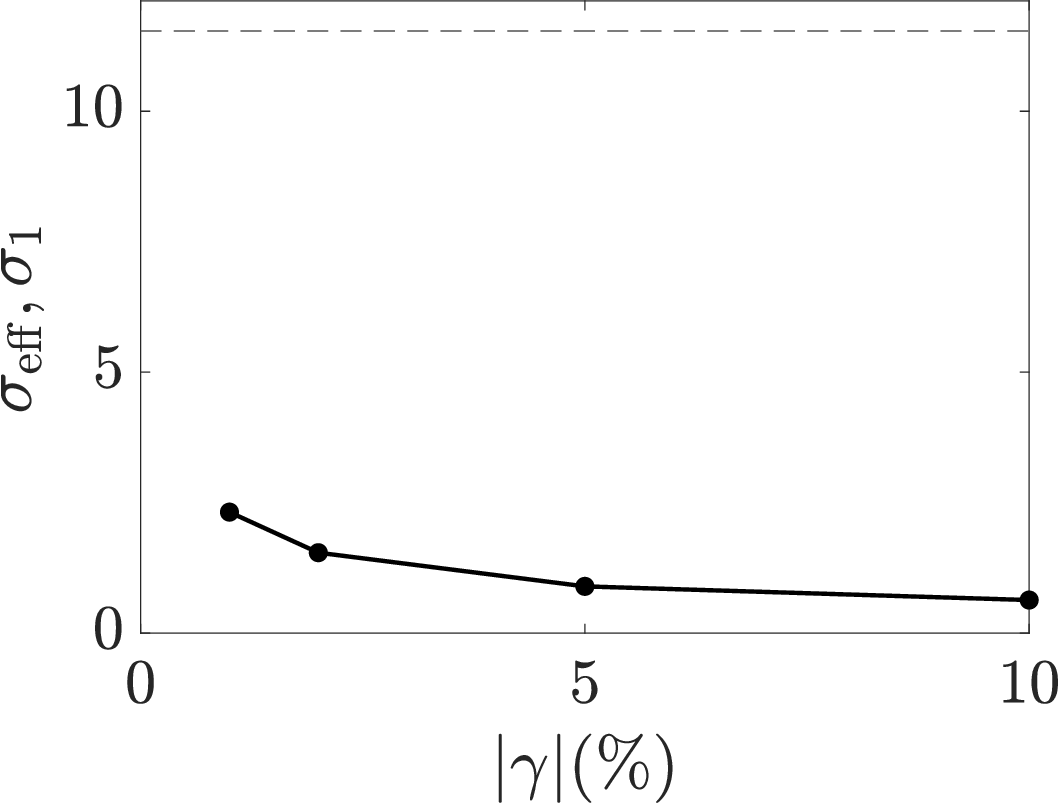}
    \caption{Effective amplification $\sigma_{\text{eff}}$ (solid) and  $\sigma_1$ (dashed).}
    \label{sigmaEffective}
\end{figure}

To measure the proportion of the linearly-amplified energy captured by the forced simulations, we compute an ensemble-averaged forcing efficiency, or effective amplification, which we define to be 
\begin{equation}
    \sigma_\text{eff} = \left(\, \frac{1}{u_\tau \delta}  \int_{0}^{T} \overline{\frac{ \Delta \hat E^{(0,1)}}{|\kappa|^2/2} } \mathrm{d}t \,\right)^{{1}/{2}}.
\end{equation}
This is analogous to $\sigma_1 = \max_{\boldsymbol{\tilde{f}}} \|\tilde{\mathcal{H}}^{(0, 1)} \mathcal B \boldsymbol{\tilde{f}} \|_2 / \|\boldsymbol{\tilde{f}}\|_2$, where the numerator reflects the energy contained in the velocity field perturbations, and the denominator corresponds to the forcing amplitude. The computed $\sigma_\text{eff}$ is shown in figure \ref{sigmaEffective}. The forcing efficiencies decrease with the intensity of the forcing, and all effective amplifications are lower than $\sigma_1 =  11.54$. This efficiency indicates that, for smaller resolvent forcing amplitudes, more of the forcing energy is linearly converted into streak energy. 

\subsection{Comparison of velocity deviations with linear response}

\begin{figure}
    \centering
    \begin{subfigure}{0.45\textwidth}
        \includegraphics[width=\linewidth]{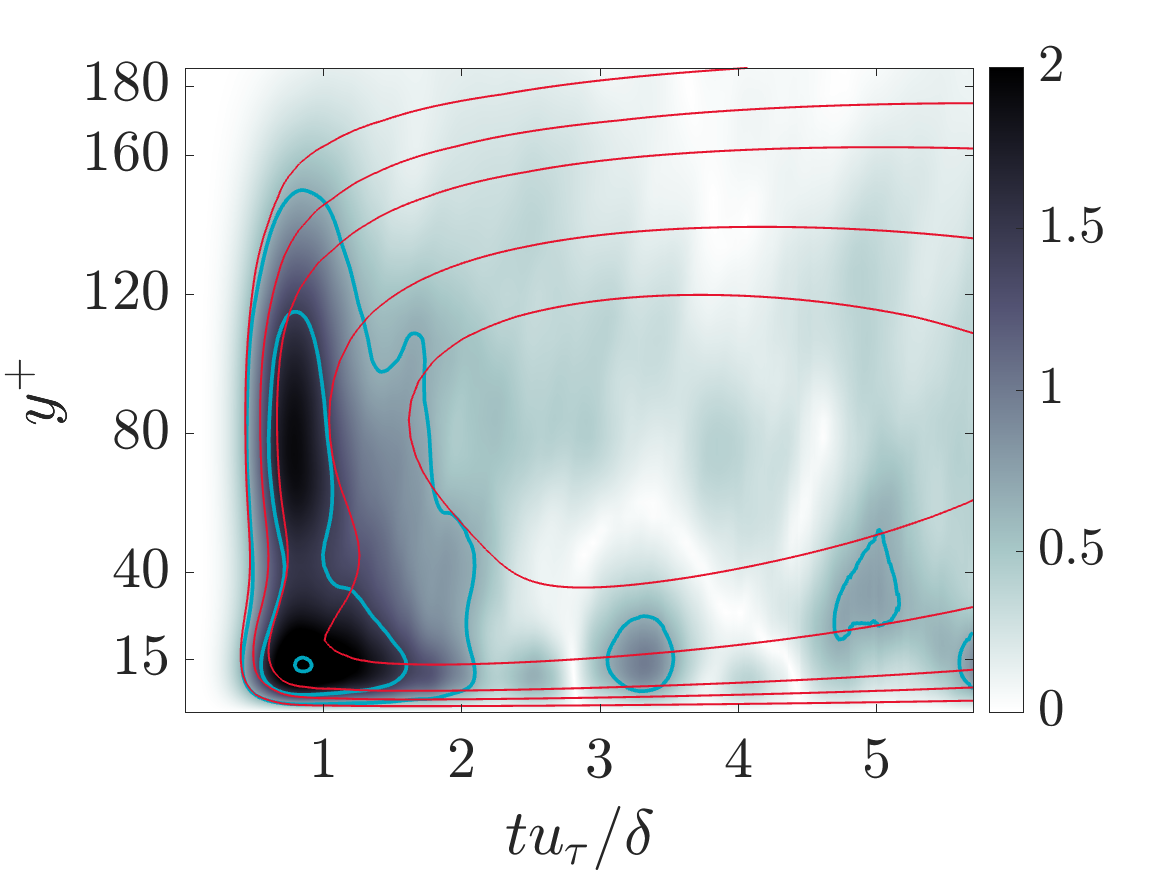}
    \end{subfigure}
    \begin{subfigure}{0.45\textwidth}
        \includegraphics[width=\linewidth]{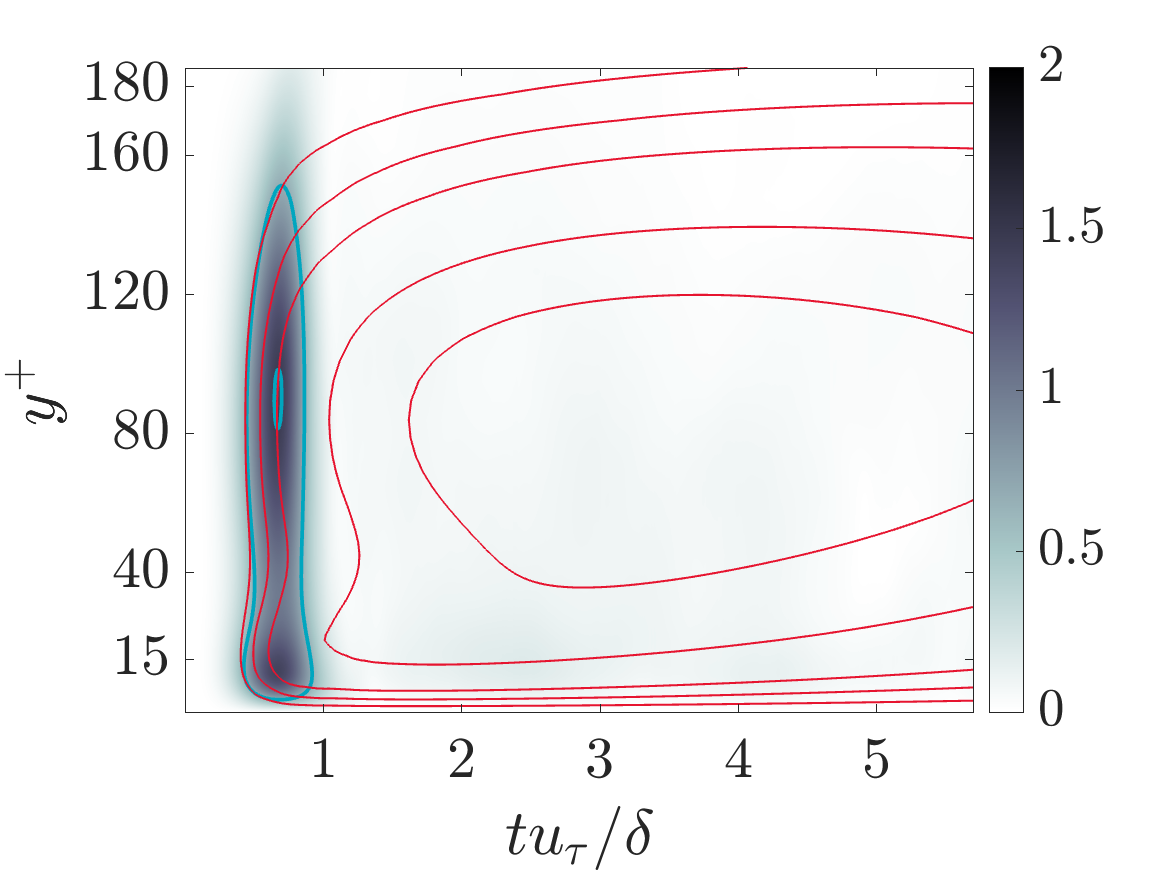}
    \end{subfigure}
    \caption{Average deviation in the streamwise velocity of the $(0,1)$-Fourier mode, $|\overline{\Delta\hat{u}^{(0,1)}/\kappa}|/u_\tau$. The contours correspond to $7\%$, $15\%$, $25\%$, $75\%$ and $90\%$ of the maximum value of $\sigma_1 \boldsymbol{\psi_1}$. The lines represent the forced DNS case (blue), and the resolvent response (red).} 
    \label{uContours}
\end{figure}

To visualize the alignment of the velocity fluctuation fields with the linear response mode across all wall normal heights, we plot the contours of the streamwise velocity magnitude deviations for the $(0, 1)$-Fourier mode, along with the contours of the linear resolvent response modes (figure \ref{uContours}). At earlier times ($t < 0.7 \delta / u_\tau $), the responses for both the $\gamma = 1\%$ and $\gamma = 10\%$ cases are very similar to the linear mode. The strongly-forced case, however, quickly reverts to the unforced channel flow statistics beyond an eddy turnover time unit. 
In contrast, the lightly-forced case of $\gamma = 1\%$ exhibits a longer-lasting velocity deviation, especially in the near wall region ($y^+ < 15$).

\begin{figure}
    \centering
    \begin{subfigure}{0.45\textwidth}
        \includegraphics[width=\linewidth]{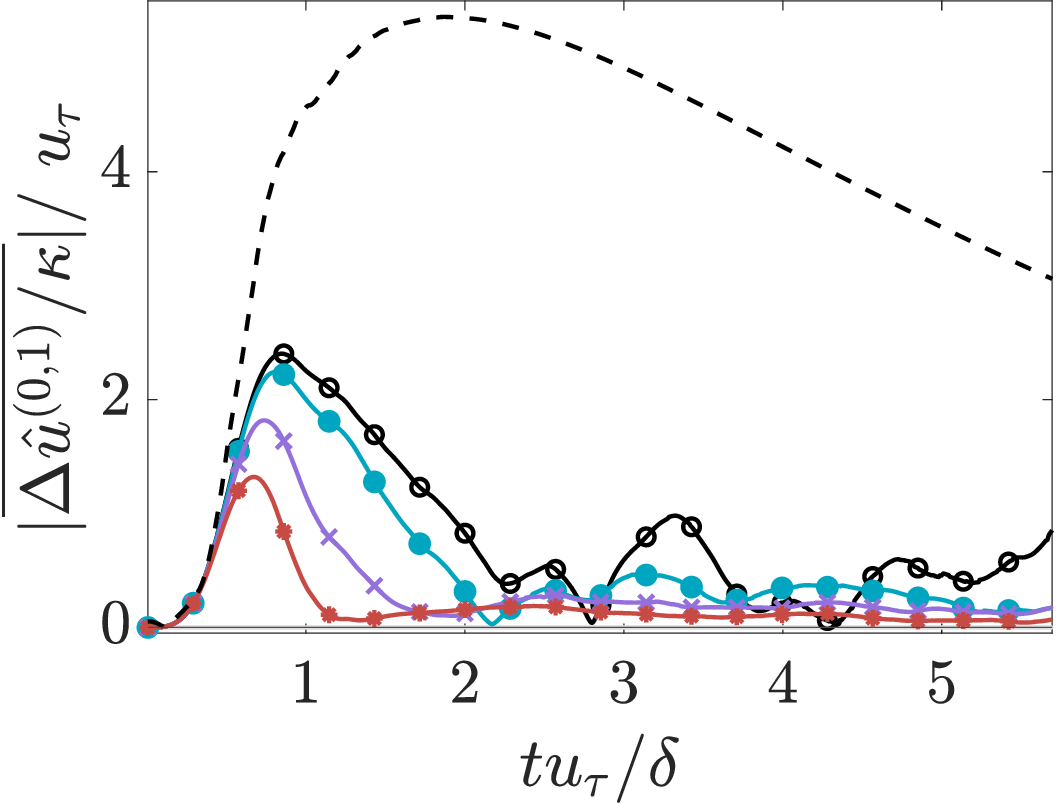}
        \caption{}
    \end{subfigure}
        \begin{subfigure}{0.45\textwidth}
        \includegraphics[width=\linewidth]{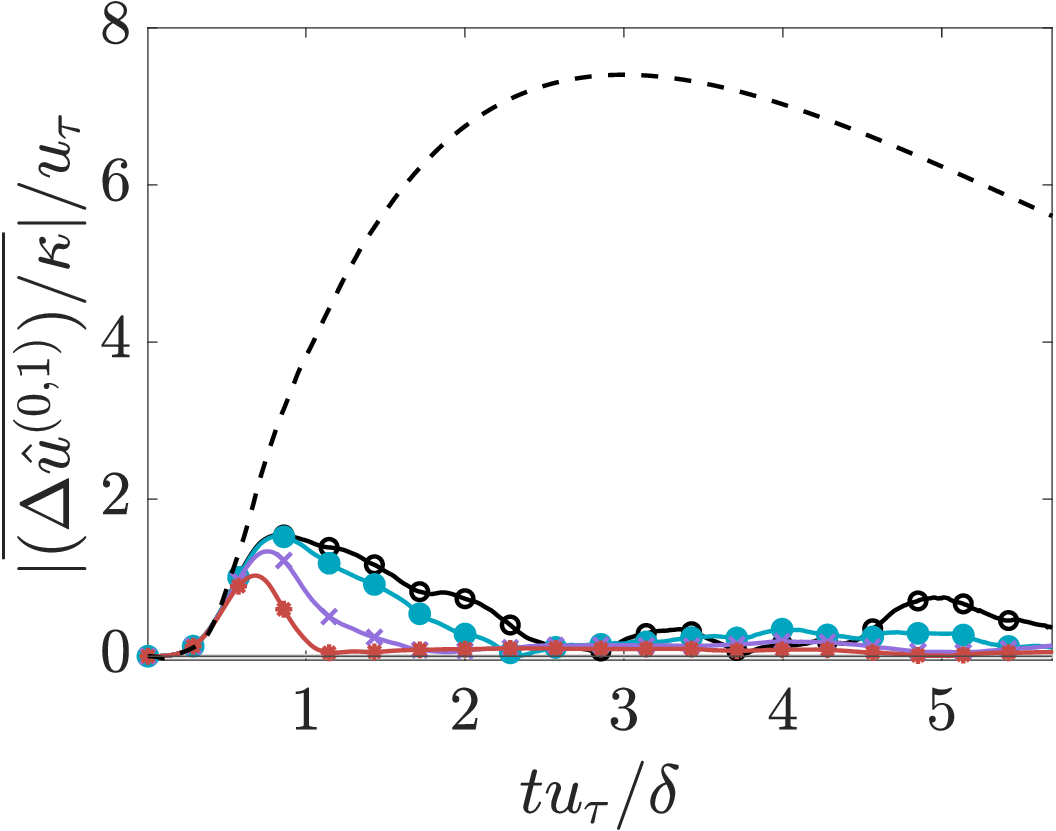}
        \caption{}
    \end{subfigure}
    \caption{
    Average deviation in the streamwise velocity of the $(0,1)$-Fourier mode at $y^+ \approx 16$ (a) and $39$ (b). The cases plotted are $\gamma = 1\%$ (black $\circ$), $\gamma = 2\%$ (cyan $\bullet$), $\gamma = 5\%$ (purple $\times$), $\gamma = 10\%$ (red $*$), and the linear response mode $\sigma_1 \boldsymbol{\psi_1}$ (black, dashed).}
    \label{absU}
\end{figure}

To more closely investigate how the agreement of the forced simulations and the optimal linear response varies with $y^+$, we show $\Delta \hat{u}^{(0,1)}$ divided by the scaling constant $\kappa$ at two wall-normal heights, along with the linear response at those heights (figure \ref{absU}).
Though the initial growth rates are similar to the linear case and we obtain good collapse prior to $t \approx 0.7 \delta /u_\tau$, the streak velocities peak earlier and decay more quickly for larger forcing amplitudes.  
For a given forcing amplitude, we notice that $\Delta \hat{u}^{(0,1)}$ diverges from the optimal linear response around the same time at both wall-normal locations plotted in figure \ref{absU}. However, as $y^+$ moves closer to the wall, the growth rate of the linear response increases, and $\Delta \hat{u}^{(0,1)}$ manages to recover more of the linearly-amplified energy before decaying, which leads to a better agreement between the forced simulations and the optimal linear response in the near-wall region. 

\begin{figure}
    \centering
    \begin{subfigure}{0.45\textwidth}
        \includegraphics[width=\linewidth]{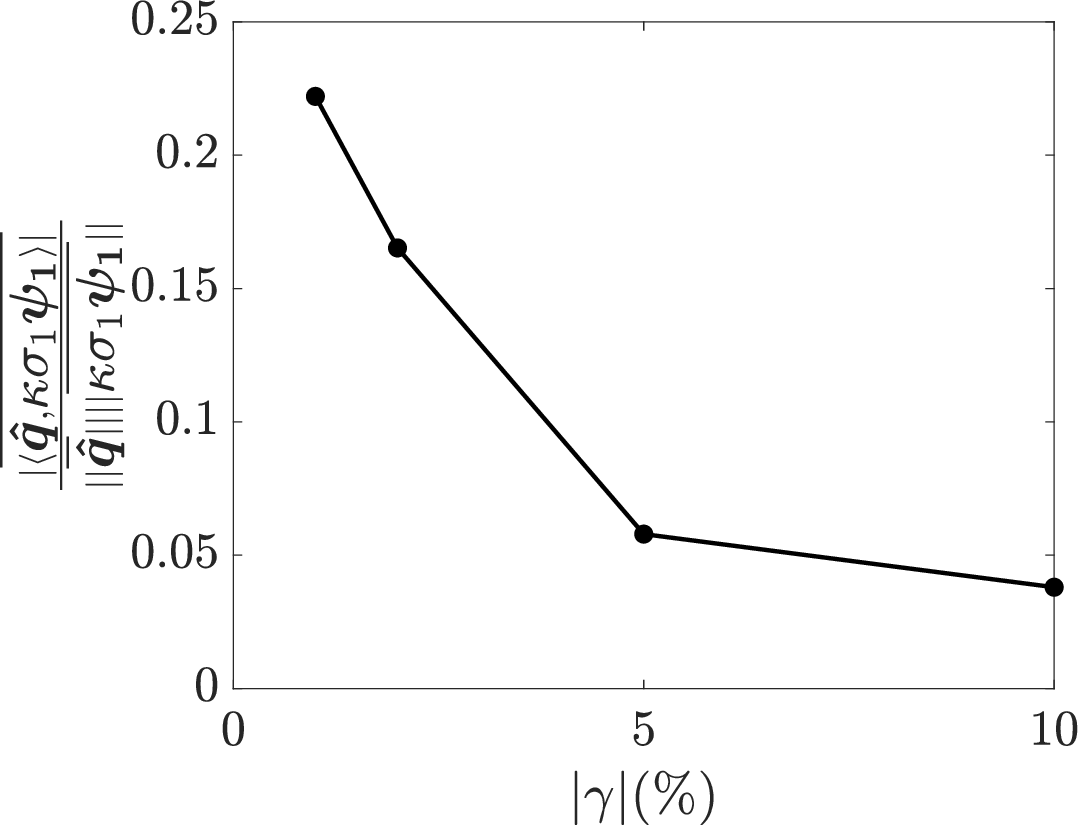}
        \caption{}
    \end{subfigure}
    \begin{subfigure}{0.45\textwidth}
        \includegraphics[width=\linewidth]{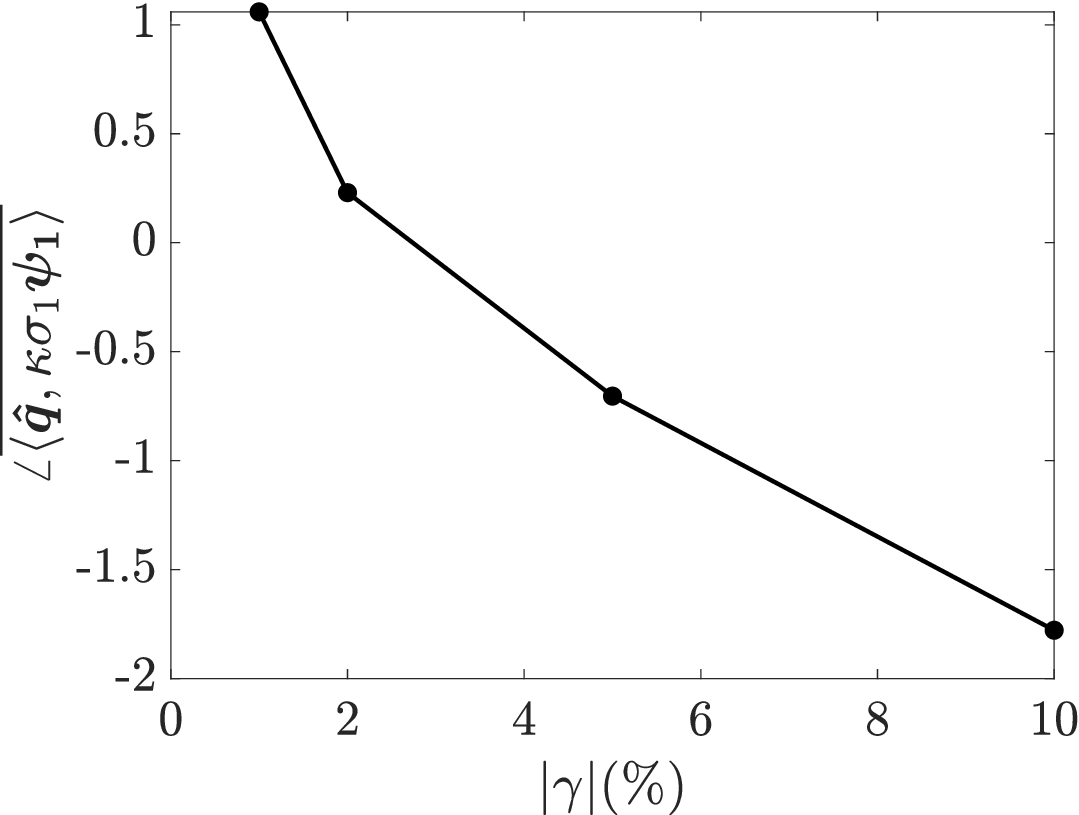}
        \caption{}
    \end{subfigure}
    \begin{subfigure}{0.45\textwidth}
        \includegraphics[width=\linewidth]{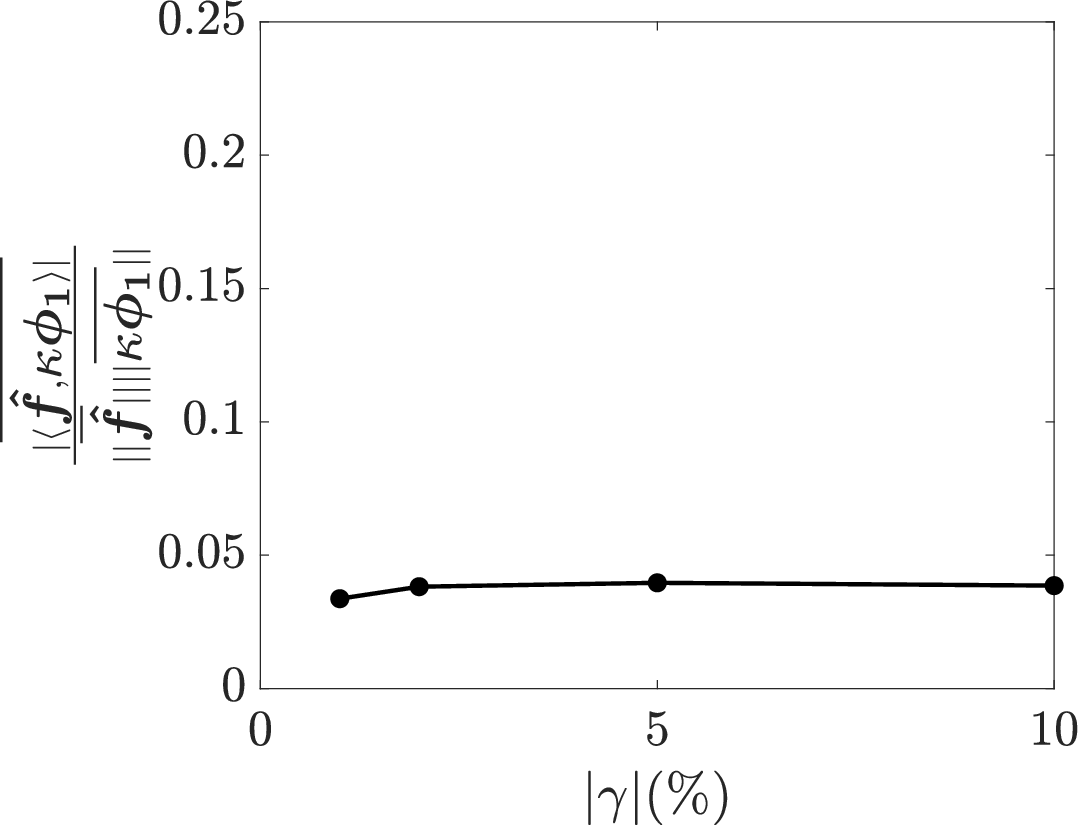}
        \caption{}
    \end{subfigure}
    \begin{subfigure}{0.45\textwidth}
        \includegraphics[width=\linewidth]{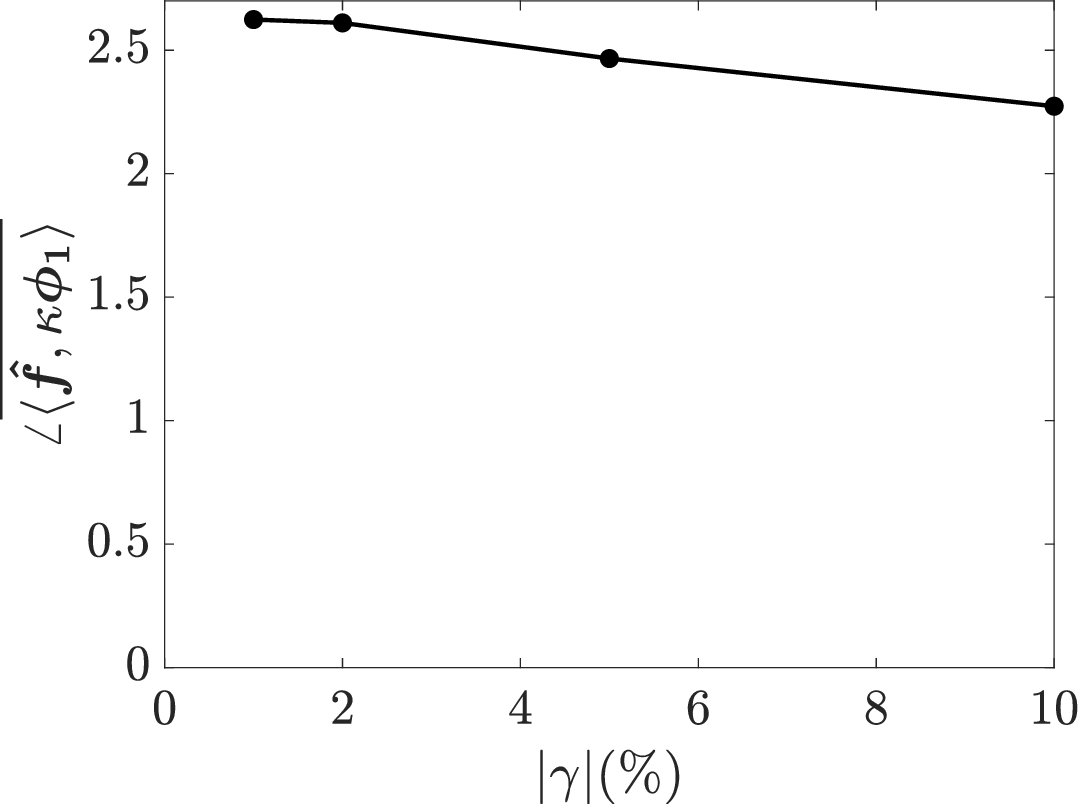}
        \caption{}
    \end{subfigure}
    \caption{Magnitude (a) and phase (b) of the average projection of the velocity field deviation onto the principle resolvent response mode $\boldsymbol{\psi_1}$. Magnitude (c) and phase (d) of the average projection of the deviation in the nonlinear terms onto the principle resolvent forcing mode $\boldsymbol{\phi_1}$.}
    \label{responseAlign}
\end{figure}

%
We quantify the linearity of the turbulent responses by projecting the deviation of the fluctuation field $\boldsymbol{\hat q}^{(0, 1)}$ onto the response mode $\kappa \sigma_1\boldsymbol{\psi_1}$. We similarly compute the projection of the deviation in nonlinear fluctuations $\boldsymbol{\hat f}^{(0, 1)}$ on the principal resolvent forcing mode $\kappa \boldsymbol{\phi_1}$. For a system governed by the linearized Navier-Stokes equations, we expect the projection of both $\boldsymbol{\hat q}^{(0, 1)}$ onto $\kappa \sigma_1\boldsymbol{\psi_1}$ and $\boldsymbol{\hat f}^{(0, 1)}$ onto $\kappa \boldsymbol{\phi_1}$ to be equal to 1. The results are shown in figure \ref{responseAlign}. 
The magnitude of the projection of $\boldsymbol{\hat q}^{(0, 1)}$ onto $\gamma\sigma_1 \boldsymbol{\psi_1}$ decreases with the intensity of the forcing (figure \ref{responseAlign}a). The magnitude trend reflects the results in figure \ref{absU}, which shows that the linear transient growth behavior lasts for shorter times as the forcing amplitude increases. The angle of the projection also varies widely for different values of $\gamma$. 

Compared to the velocity fluctuation projection, the magnitude of the projection of the nonlinear term is much lower and relatively constant with $\gamma$. This shows that, across all forcing cases, the injected resolvent mode accounts for very little of the Reynolds stresses in the system. Considering this result, it is remarkable that the forcing successfully produces significant energy growth that tracks the optimal linear response to the observed degree.

\begin{figure}
    \centering
    \begin{subfigure}{0.31\textwidth}
        \includegraphics[width=\linewidth]{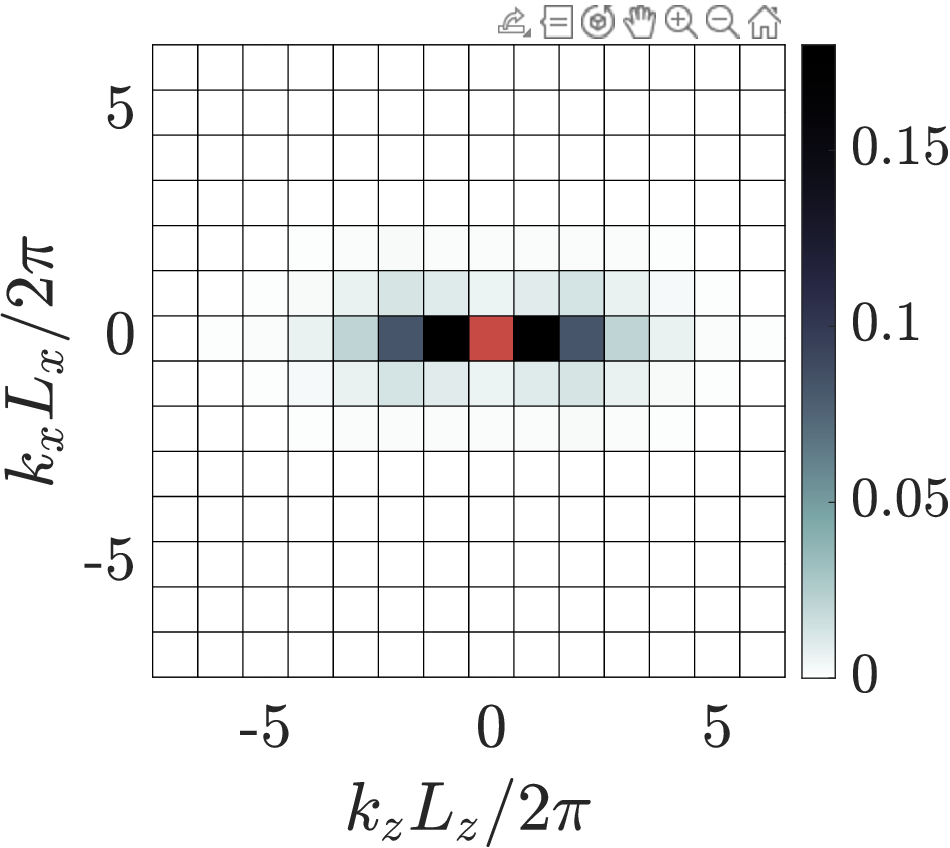}
    \end{subfigure}
    \begin{subfigure}{0.31\textwidth}
        \includegraphics[width=\linewidth]{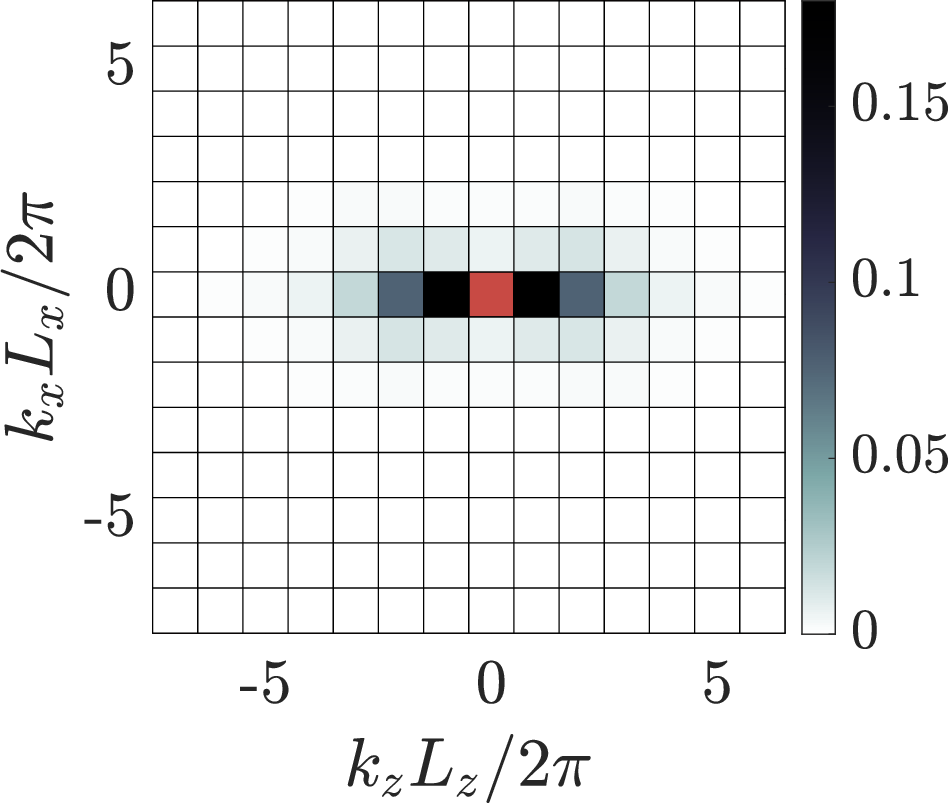}`
    \end{subfigure}
    \begin{subfigure}{0.31\textwidth}
        \includegraphics[width=\linewidth]{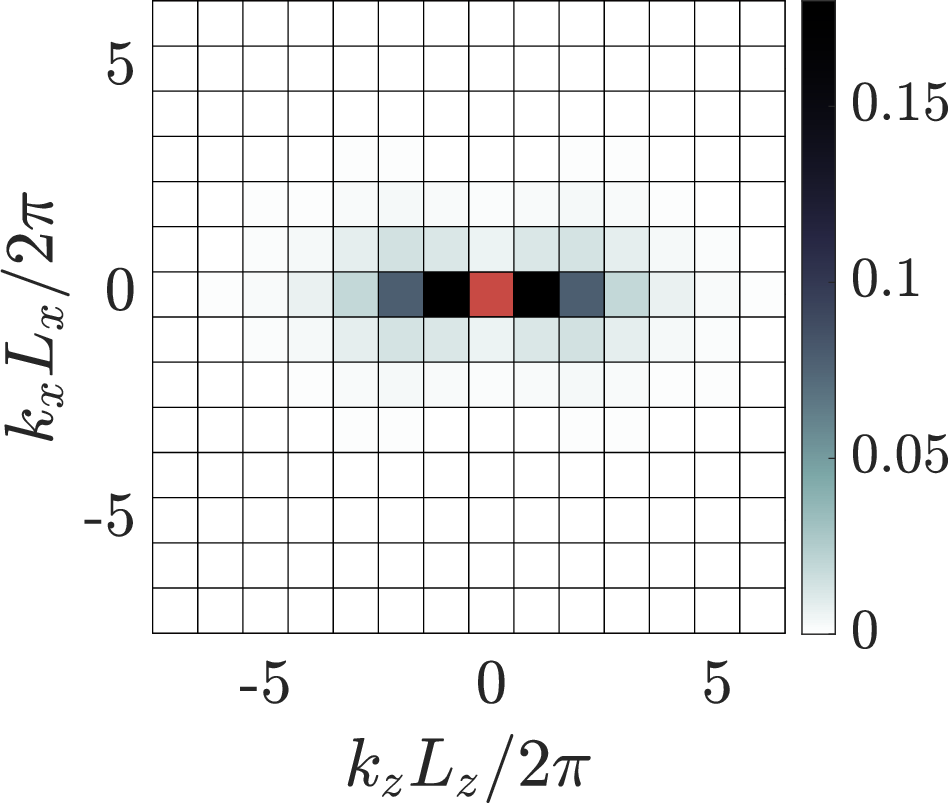}`
    \end{subfigure}
    \begin{subfigure}{0.31\textwidth}
        \includegraphics[width=\linewidth]{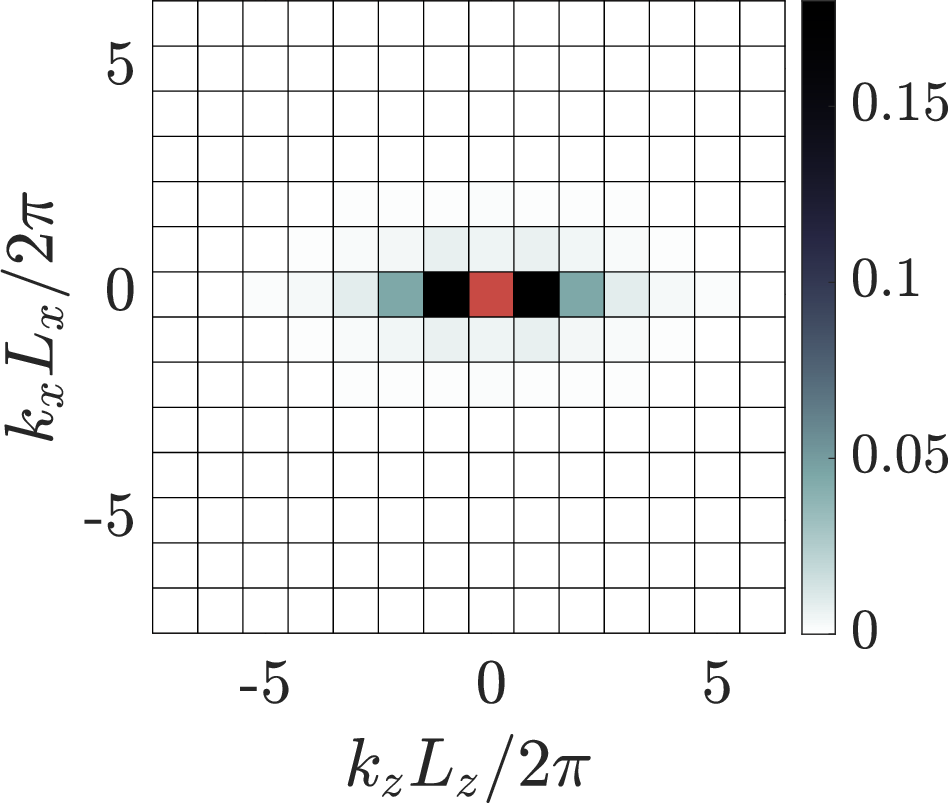}
    \end{subfigure}
    \begin{subfigure}{0.31\textwidth}
        \includegraphics[width=\linewidth]{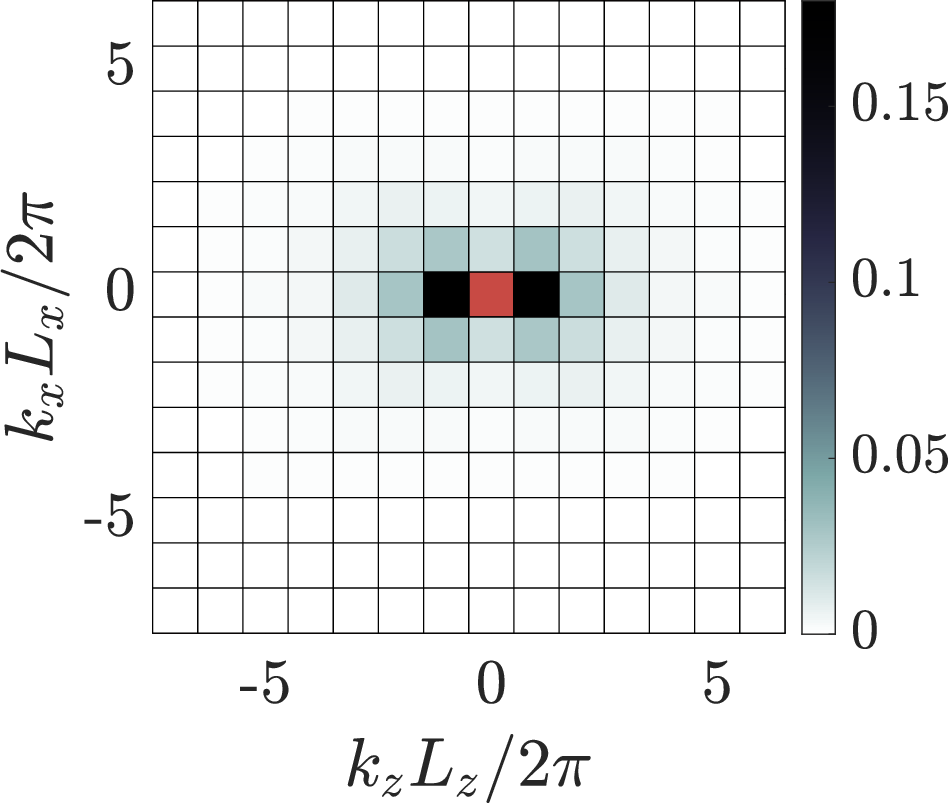}
    \end{subfigure}
    \begin{subfigure}{0.31\textwidth}
        \includegraphics[width=\linewidth]{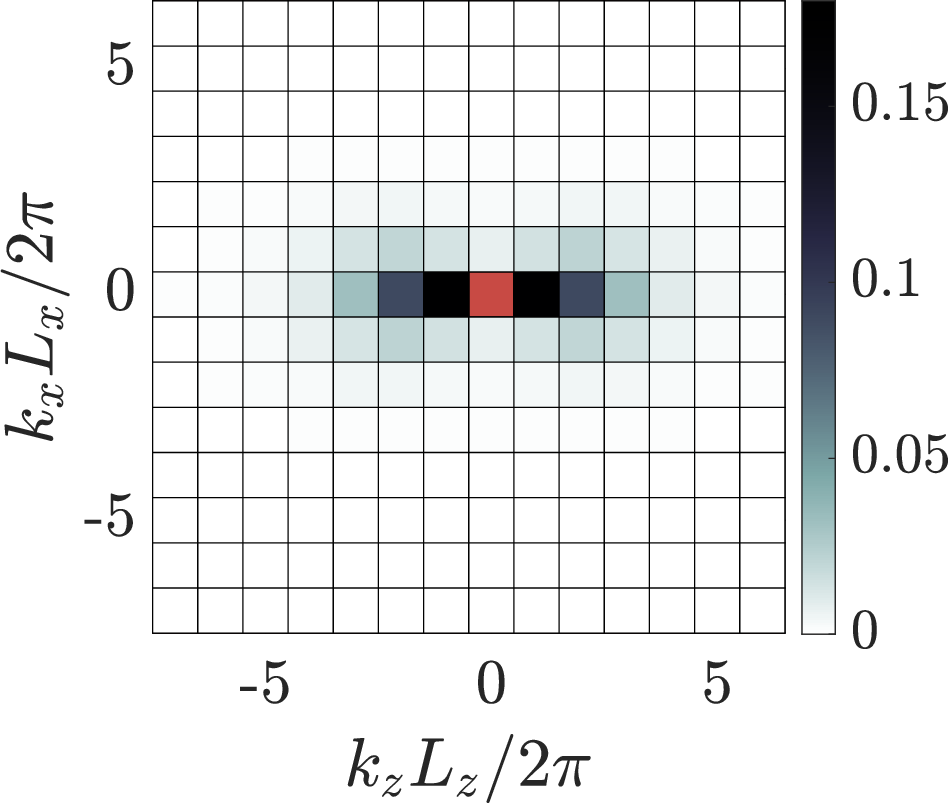}
    \end{subfigure}
    \caption{Streamwise spectral energy content $\overline{|\hat{u}|^2}/(2  u_\tau^2)$ at $y^+ \approx 16$ for $\gamma = 1\%$ (top) and $\gamma = 10\%$ (bottom). The spectra shown are at times $t = 0.5 \delta/u_\tau$ (left), $1 \delta/u_\tau$ (middle) and $2 \delta/u_\tau$ (right). The energy contained in the $(0,0)$--mode (red) is excluded for clarity.}
    \label{SpectrumFixedTimes}
\end{figure}

\begin{figure}
    \centering
    \begin{subfigure}{0.45\textwidth}
    \includegraphics[width=\linewidth]{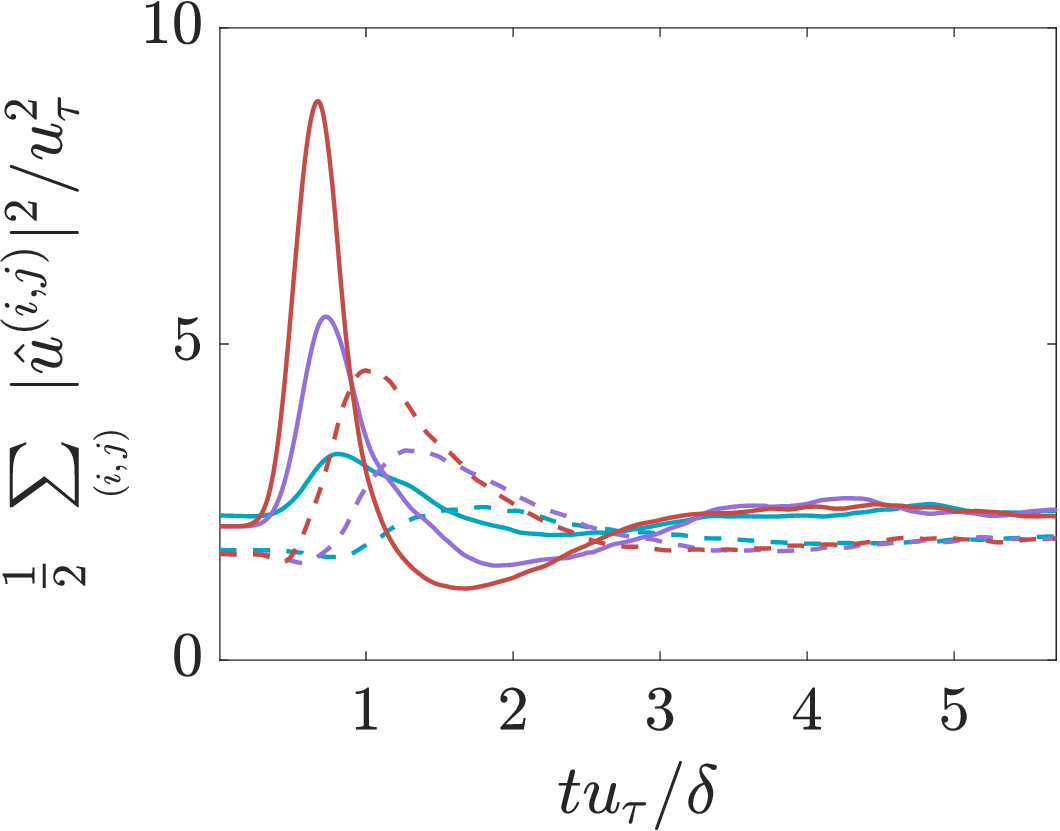}
    \caption{}
    \end{subfigure}
    \begin{subfigure}{0.45\textwidth}
    \includegraphics[width=\linewidth]{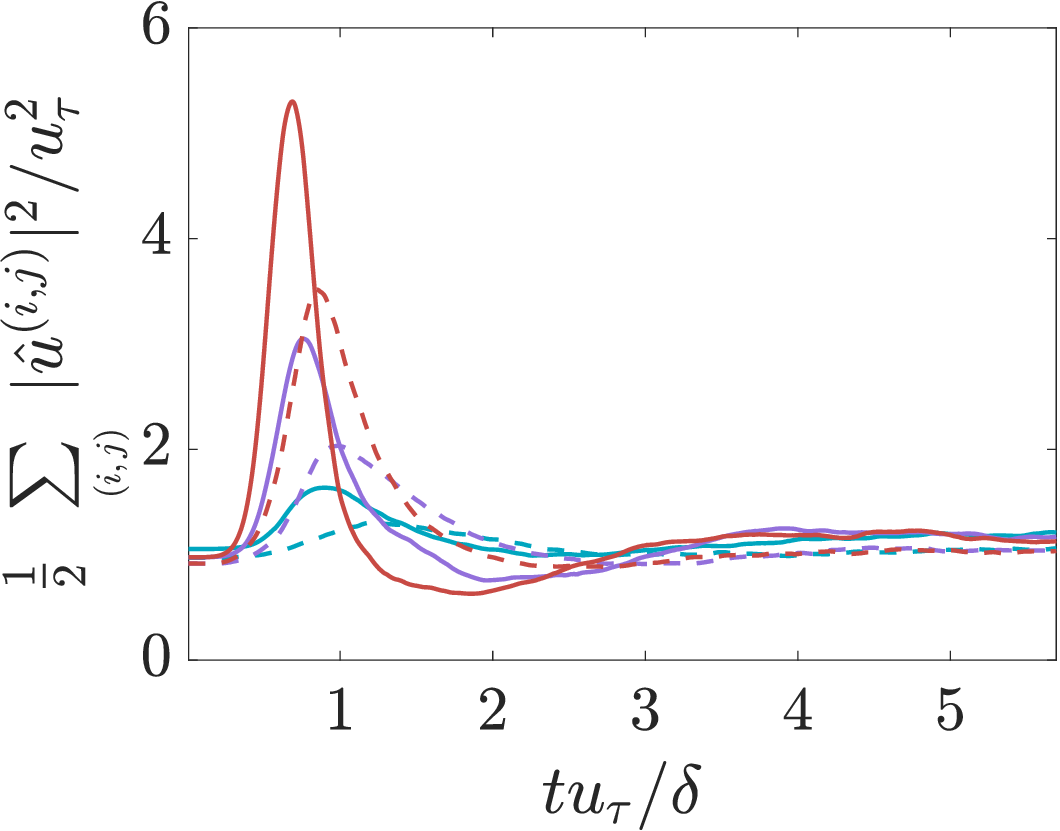}
    \caption{}
    \end{subfigure}
    \begin{subfigure}{0.44\textwidth}
    \includegraphics[width=\linewidth]{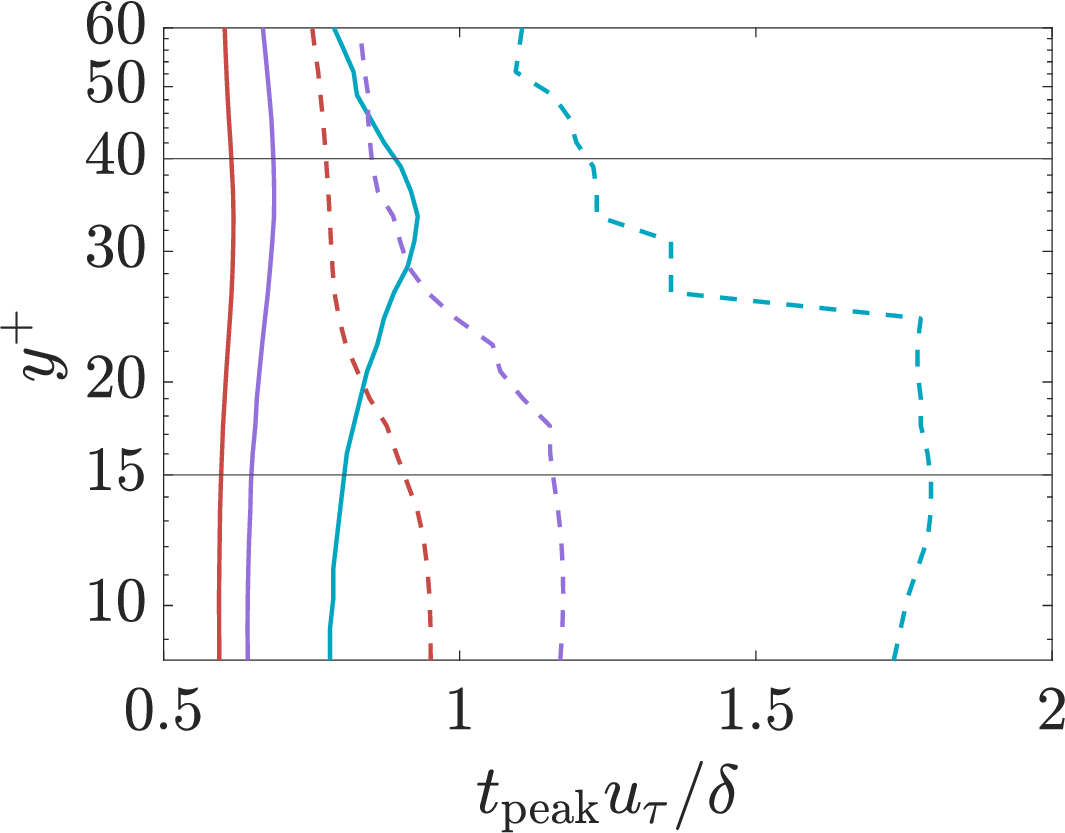}
    \caption{}
    \end{subfigure}
    \begin{subfigure}{0.44\textwidth}
    \includegraphics[width=\linewidth]{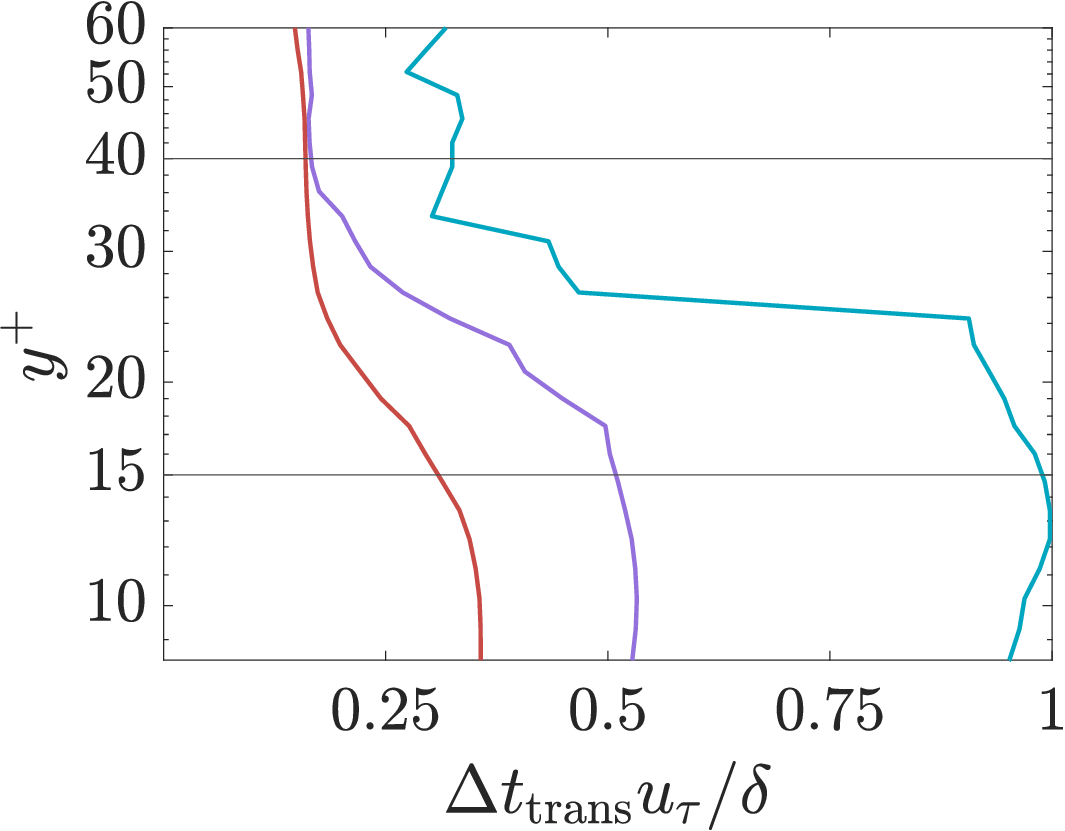}
    \caption{}
    \end{subfigure}
    
    \caption{Streamwise spectral energy content at $y^+ \approx 16$ (a) $y^+ \approx 39$ (b) and  for the $(0,\pm 1)$--Fourier modes (solid lines) and the other fluctuation wavenumbers (dashed lines). 
    (c) Energy peak times for the $(0, \pm1)$--mode (solid) and the smaller scales (dashed) cases. 
    (d) Timescale for the cross-scale energy transport. Colors indicate $\gamma = 2\%$  (blue), $\gamma \approx 5\%$ (purple), and $\gamma = 10\%$ (red).  The horizontal black lines in (c) and (d) represent $y^+ = 15$ and $y^+ = 40$, which delineate the buffer layer.}
    \label{01mode}
\end{figure}

\subsection{Spectra}

We compute the streamwise and spanwise spectra at a wall-normal height of $y^+ \approx 16$ for $\gamma = 1\%$ and $\gamma = 10\%$, before the energy peak ($t = 0.5 \, \delta/ u_\tau$), during energy decay ($t = \delta/ u_\tau$), and after decay ($t = 2 \, \delta/ u_\tau$). The results are shown figure \ref{SpectrumFixedTimes}. Across all times, the $(0, \pm 1)$--mode is the most dominant one, accounting for $30\% - 70\%$ of the total turbulent energy, followed by the $(0, \pm 2)$-- mode accounting for $6\% - 18\%$. During the energy decay of the $(0,1)$--mode, the strongly-forced case exhibits energy transfer to secondary modes, especially the $(\pm 1, \pm1)$--modes. These multiscale effects are not as prominent for the lightly-forced case. 

The energy cascade to smaller scales is clearly visible in figures \ref{01mode}a and \ref{01mode}b, which plot the energy content in the $(0, \pm 1)$--mode and the smaller scales as function of time for the different forcing amplitude cases at $y^+ \approx 16$ and $y^+ \approx 39$. Across all forcing amplitudes, the rise of the energy content of smaller scales coincides with the start of the energy decay of the $(0,1)$--mode, \emph{i.e.}, the interrupted growth and faster decrease of $ |\overline{\hat u^{(0,1)}}|^2/(2u_\tau^2)$ compared to the optimal linear response.  
At both wall normal heights shown in figures \ref{01mode}a and \ref{01mode}b, we observe that a higher amplification triggers an earlier and larger energy leak into the smaller spatial scales. Furthermore, for a given forcing amplitude, the energy content of the small scales rises faster and peaks earlier farther away from the wall. 

To better visualize the cross-scale energy transport timescales, we plot the streamwise energy peak times for the $(0, \pm1)$--mode and the smaller scales as a function of $y^+$ and for different forcing amplitudes (figures \ref{01mode}c). 
For all wall-normal heights, increasing the forcing amplitude indeed causes the energy of both the $(0, \pm1)$--mode and the smaller scales to peak earlier. 
For a given forcing amplitude, the peak time for the $(0, \pm1)$--mode is constant with $y^+$ in the near wall region, increases slightly with $y^+$ in the buffer layer, and plateaus again in the outer region of the flow. 
The variation in the energy peak times for the smaller scales is more dramatic: the peak time is also constant in the near-wall region, but decreases significantly with $y^+$ within the buffer layer, and levels off in the outer region. 

We define the timescale for cross-scale energy transport, $\Delta t_\mathrm{trans}$, as the time delay between the energy peaks of the $(0, \pm1)$--modes and the smaller scales; 
its dependence on $y^+$ is mostly determined by the energy peak time for the smaller scales (figure \ref{01mode}d).
Interestingly, though a larger forcing amplitudes accelerates the energy transport from the $(0, \pm 1)$--mode to smaller scales for all wall normal heights, the sensitivity of $\Delta t_\mathrm{trans}$ to forcing amplitude decreases as $y^+$ moves farther away from the wall. Indeed, in the outer regions of the flow, $\Delta t_\mathrm{trans}$ converges to a value of approximately $0.18 \, \delta / u_\tau $ for high forcing amplitude.

\begin{figure}
    \centering
    \includegraphics[width=0.45\textwidth]{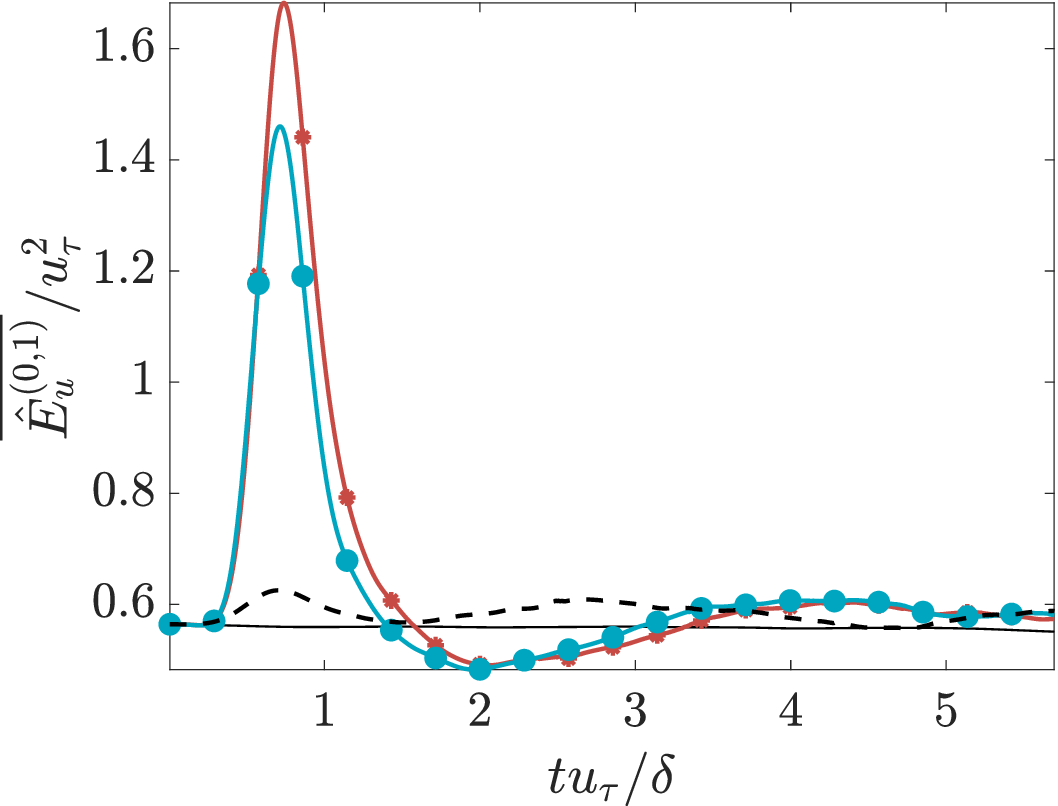}
    \caption{Average streak energy for forcing terms aligned with $\boldsymbol{\phi_1}$ (red $*$), $\boldsymbol{\phi_3}$ (cyan $\bullet$), and $\boldsymbol{\phi}_\mathrm{\bf rand}$ (dashed black). The unforced case is shown in black.}
    \label{optimality}
\end{figure}

\subsection{Optimality of resolvent forcing}
We finally compare the results of forcing using the principal resolvent forcing mode $\boldsymbol{\phi_1}$, the first suboptimal mode $\boldsymbol{\phi_3}$, and the random mode $\boldsymbol{\phi}_{\text{\bf{rand}}}$.
We see that the streak energy grows to higher peak when the minimal flow unit is forced by $\boldsymbol{\phi_1}$ than when forced by $\boldsymbol{\phi_3}$ (figure \ref{optimality}). In both cases, the minimal flow unit is much more responsive compared to the case with random forcing. This suggests that resolvent analysis identifies a forcing structure to which the minimal flow unit is indeed sensitive, even when governed by the fully nonlinear Navier-Stokes equations. 
We note, however, that the advantage the optimal forcing mode is significantly reduced as the effective amplifications, $\sigma_\text{eff}$, in DNS forced by $\boldsymbol{\phi_1}$ and $\boldsymbol{\phi_3}$ differ only by a factor of $1.03$, whereas $\sigma_1/\sigma_3 = 2.16$.


\section{Conclusions}\label{conclusion}

In this work, we studied the growth of time-localized resolvent modes in the minimal flow unit at $Re_\tau \approx 186$. We formulated resolvent analysis in a wavelet-basis in time that endows the resolvent modes with transient information, and obtained a linearly optimal time-localized forcing mode and its corresponding transient response mode.

Resolvent analysis ignores the feedback between the velocity fluctuations and the nonlinear terms; we thus tested the optimality of the resolvent forcing within a DNS of a minimal flow unit by numerically injecting the principal resolvent forcing mode into the flow at varying amplitudes. This allowed us to investigate the interactions between the transiently growing linear response mode and the nonlinear effects of the turbulent flow.
We compared the resulting flow to one forced by the first suboptimal forcing mode and another forced by a spatially-random forcing. The principal resolvent forcing produces a larger transient energy growth than the suboptimal mode, but the energy amplification was notably lower in both cases compared to the linearized case. Both systems were significantly more amplified than the random forcing case. In all cases, the injected forcing term is negligible compared to the newly-induced Reynolds stresses, and it is remarkable that it still manages to produce a significant energy amplification and a perturbation to the velocity field that tracks the optimal linear response for a short time.

The nonlinearities of turbulence interrupt the initial algebraic energy growth driven by the linear dynamics of the flow. This is seen in all cases forced by the principal resolvent mode, though the amplitude of the resolvent forcing affects how closely the turbulent trajectory behaves like the optimal resolvent response mode. 
Across all forcing amplitudes, the initial growth phase is similar, and the systems peak at roughly the same time. However, the higher the forcing amplitude, the faster the decay of the system back to the unforced turbulent system. The more intense Reynolds stresses in the high-amplitude-forcing cases are more effective at damping the effects of the initial forcing. 
Additionally, the forced DNS flow fields are closer to the resolvent response mode in the near-wall region. 
The spectral energy content becomes increasingly nonlinear and multi-scale during the decay phase due to a transfer of energy to the non-forced spatial scales. This cross-scale energy transfer is more prominent and occurs faster for the high-amplitude-forcing cases.
The nonlinear effects interrupt the initial linear energy growth mechanism and lead to streak breakdown. 

\section{Acknowledgments}
This work was supported in part by the European Research Council under the Caust grant ERC-AdG-101018287 and AFOSR grant FA9550-22-1-0109.
\section*{References}
\bibliography{references}

\providecommand{\newblock}{}
\begin{thebibliography}{10}
\expandafter\ifx\csname url\endcsname\relax
  \def\url#1{{\tt #1}}\fi
\expandafter\ifx\csname urlprefix\endcsname\relax\def\urlprefix{URL }\fi
\providecommand{\eprint}[2][]{\url{#2}}

\bibitem{klebanoff1962three}
Klebanoff P~S, Tidstrom K~D and Sargent L~M 1962 {\em J. Fluid Mech.\/} {\bf 12} 1--34

\bibitem{kline1967structure}
Kline S~J, Reynolds W~C, Schraub F~A and Runstadler P~W 1967 {\em J. Fluid Mech.\/} {\bf 30} 741--773

\bibitem{smithmetzler1983characteristics}
Smith C and Metzler S 1983 {\em J. Fluid Mech.\/} {\bf 129} 27--54

\bibitem{blackwelder1979streamwise}
Blackwelder R~F and Eckelmann H 1979 {\em J. Fluid Mech.\/} {\bf 94} 577--594

\bibitem{johansson1987generation}
Johansson A~V, Her J~Y and Haritonidis J~H 1987 {\em J. Fluid Mech.\/} {\bf 175} 119--142

\bibitem{bakewell1967viscous}
Bakewell~Jr H~P and Lumley J~L 1967 {\em Phys. Fluids\/} {\bf 10} 1880--1889

\bibitem{landahl1980note}
Landahl M 1980 {\em J. Fluid Mech.\/} {\bf 98} 243--251

\bibitem{butler1993optimal}
Butler K~M and Farrell B~F 1993 {\em Phys. Fluids A\/} {\bf 5} 774--777

\bibitem{chernyshenko2005mechanism}
Chernyshenko S and Baig M 2005 {\em J. Fluid Mech.\/} {\bf 544} 99--131

\bibitem{delalamo2006linear}
Del~Alamo J~C and Jimenez J 2006 {\em J. Fluid Mech.\/} {\bf 559} 205--213

\bibitem{kim1971production}
Kim H~T, Kline S and Reynolds W~C 1971 {\em J. Fluid Mech.\/} {\bf 50} 133--160

\bibitem{hamilton1995regeneration}
Hamilton J~M, Kim J and Waleffe F 1995 {\em J. Fluid Mech.\/} {\bf 287} 317--348

\bibitem{panton2001overview}
Panton R~L 2001 {\em Prog. Aerosp. Sci.\/} {\bf 37} 341--383

\bibitem{jimenez2018coherent}
Jim{\'e}nez J 2018 {\em J. Fluid Mech.\/} {\bf 842} P1

\bibitem{smits2011high}
Smits A~J, McKeon B~J and Marusic I 2011 {\em Annu. Rev. Fluid Mech.\/} {\bf 43} 353--375

\bibitem{adrian2007hairpin}
Adrian R~J 2007 {\em Phys. Fluids\/} {\bf 19}

\bibitem{robinson1991coherent}
Robinson S~K 1991 {\em Annu. Rev. Fluid Mech.\/} {\bf 23} 601--639

\bibitem{jimenez1991minimal}
Jim{\'e}nez J and Moin P 1991 {\em J. Fluid Mech.\/} {\bf 225} 213--240

\bibitem{jimenez1999autonomous}
Jim{\'e}nez J and Pinelli A 1999 {\em J. Fluid Mech.\/} {\bf 389} 335--359

\bibitem{lozano2021cause}
Lozano-Dur{\'a}n A, Constantinou N~C, Nikolaidis M~A and Karp M 2021 {\em J. Fluid Mech.\/} {\bf 914} A8

\bibitem{jimenez2013linear}
Jim{\'e}nez J 2013 {\em Phys. Fluids\/} {\bf 25}

\bibitem{orr1907stability}
Orr W~M~F 1907 {\em Proc. R. Ir. Acad.\/} vol~27 (JSTOR) pp 69--138

\bibitem{hwang2010self}
Hwang Y and Cossu C 2010 {\em Phys. Rev. Lett.\/} {\bf 105} 044505

\bibitem{pujals2009note}
Pujals G, Garc{\'\i}a-Villalba M, Cossu C and Depardon S 2009 {\em Phys. Fluids\/} {\bf 21} 015109

\bibitem{mckeon2010critical}
McKeon B~J and Sharma A~S 2010 {\em J. Fluid Mech.\/} {\bf 658} 336--382

\bibitem{moarref2014foundation}
Moarref R, Sharma A~S, Tropp J~A and McKeon B~J 2013 {\em J. Fluid Mech.\/} {\bf 734} 275--316

\bibitem{mckeon2017engine}
McKeon B 2017 {\em J. Fluid Mech.\/} {\bf 817} P1

\bibitem{bae2021nonlinear}
Bae H~J, Lozano-Duran A and McKeon B~J 2021 {\em J. Fluid Mech.\/} {\bf 914} A3

\bibitem{bae2018}
Bae H~J, Lozano-Dur\'an A, Bose S~T and Moin P 2018 {\em Phys. Rev. Fluids\/} {\bf 3}(1) 014610

\bibitem{ballouz2023wavelet}
Ballouz E, Lopez-Doriga B, Dawson S~T and Bae H~J 2023 {\em AIAA SCITECH 2023 Forum\/} p 0676

\bibitem{orlandi2000fluid}
Orlandi P 2000 {\em Fluid flow phenomena: a numerical toolkit\/} vol~55 (Springer Science \& Business Media)

\bibitem{kim1985application}
Kim J and Moin P 1985 {\em J. Comput. Phys.\/} {\bf 59} 308--323

\bibitem{wray1990minimal}
Wray A~A 1990 {\em NASA Ames Research Center, California, Report No. MS\/} {\bf 202}

\bibitem{lozano2016turbulent}
Lozano-Dur{\'a}n A and Bae H 2016 {\em Annual Research Briefs. Center for Turbulence Research\/} {\bf 2016} 97--103

\bibitem{bae2018turbulence}
Bae H~J, Lozano-Duran A, Bose S and Moin P 2018 {\em Phys. Rev. Fluids\/} {\bf 3} 014610

\bibitem{bae2019dynamic}
Bae H~J, Lozano-Dur{\'a}n A, Bose S~T and Moin P 2019 {\em J. Fluid Mech.\/} {\bf 859} 400--432

\bibitem{mallat2001wavelet}
Mallat S 1999 {\em A wavelet tour of signal processing\/} (Elsevier)

\bibitem{najmi2012wavelets}
Najmi A~H 2012 {\em Wavelets: A concise guide\/} (JHU Press)

\bibitem{jeun2016input}
Jeun J, Nichols J~W and Jovanovi{\'c} M~R 2016 {\em Phys. Fluids\/} {\bf 28} 047101

\bibitem{kojima2020}
Kojima Y, Yeh C, Taira K and Kameda M 2020 {\em J. Fluid Mech.\/} {\bf 885} R1

\bibitem{daubechies1992ten}
Daubechies I 1992 {\em Ten lectures on wavelets\/} (SIAM)

\end{thebibliography}

\end{document}